\documentclass[prl,showpacs,twocolumn,superscriptaddress]{revtex4-1}
\usepackage{amsmath,amssymb,mathrsfs}
\usepackage[dvips]{graphicx}
\usepackage{hyperref}
\usepackage{paralist}

\newcommand{\e}{{\rm e}}
\newcommand{\el}[1]{\label{#1}}
\newcommand{\fl}[1]{\label{#1}}
\newcommand{\eref}[1]{\ref{#1}}
\newcommand{\fref}[1]{\ref{#1}}
\newcommand{\zd}{{\rm d}}
\usepackage{dsfont}
\allowdisplaybreaks[4]

\def\be{\begin{equation}}
\def\ee{\end{equation}}
\def\bea{\begin{eqnarray}}
\def\eea{\end{eqnarray}}

\begin{document}

\title{Understanding many-body physics in one dimension  from the Lieb-Liniger model }

\author{Y.-Z. Jiang}
\author{Y.-Y. Chen}
\author{X.-W. Guan}
\email[e-mail:]{xwe105@wipm.ac.cn}

\affiliation{State Key Laboratory of Magnetic Resonance and Atomic and Molecular Physics,
Wuhan Institute of Physics and Mathematics, Chinese Academy of Sciences, Wuhan 430071, China}




\affiliation{Center for Cold Atom Physics, Chinese Academy of Sciences, Wuhan 430071, China}

\date{\today}

\begin{abstract}
This article presents an elementary  introduction on  various aspects of the  prototypical integrable model the  Lieb-Liniger Bose gas ranging from the cooperative to the collective features of many-body phenomena.
In 1963 Lieb and Liniger   first  solved this quantum field theory many-body problem   using  the  Bethe's  hypothesis, i.e.  a particular form of wave function   introduced by Bethe in solving the one-dimensional Heisenberg model  in 1931.
Despite  the Lieb-Liniger  model  is  arguably   the simplest exactly solvable model, it exhibits  rich  quantum many-body physics in terms of  the aspects of  mathematical integrability and physical universality.
Moreover, the Yang-Yang grand canonical ensemble description for the model provides us with a deep understanding of quantum statistics, thermodynamics and quantum critical phenomena  at the many-body physics  level.
Recently,  such  fundamental physics of this exactly solved  model has been attracting  growing interest  in experiments. 
Since 2004, there have been more than  20 experimental papers that report novel observations of   different physical aspects of the Lieb-Liniger  model  in the lab.
So far the observed results to date are seen to be in excellent agreement with results obtained using the analysis of this simplest exactly solved model.
Those experimental observations   reveal  the unique  beauty of integrability.

\end{abstract}


\maketitle

\section{Introduction}

Mathematical principles play significant  roles in understanding quantum physics. The concept ``exact integrability'' originally came from the  early study  of  the  classical dynamical systems  which were described by some differential equations.
Usually, the solutions of those differential equations are tantamount to the determination of enough integration constants, i.e., the integrals of motions.
In this sense, classical integrability is synonymous with exact solution.
However, conceptual understanding of quantum integrability should trace back to Hans Bethe's seminal work to obtain the energy eigenstates of the one-dimensional Heisenberg spin chain with the nearest interaction in 1931.\cite{1}
He proposed a special form of the wavefunction --- superposition of all possible permutations of plane waves in a ring of size $L$, namely,
\begin{eqnarray}
 \chi &=&\sum_{\cal P}A({\cal P}) {\rm e}^{\mathrm{i}(k_{{\cal P}_1}x_{1}+\cdots+k_{{\cal P}_N}x_{N})},\nonumber
\end{eqnarray}
where $N$ is the number of down spins and ${\cal P}_1,\ldots, {\cal P}_N$ stand for a permutation ${\cal P}$ of $1,\,2, \, \ldots,\,N$.
The $N!$ plane waves are $N$-fold products of individual exponential phase
factors $\e^{\mathrm{i}k_ix_j}$.
Here the $N$ distinct wave numbers $k_i$ are permuted among the $N$ distinct coordinates $x_j$.
Each of the $N!$ plane waves has an amplitude coefficient which can be in turn determined by solving the eigenvalue problem of the Heisenberg Hamiltonian.

The Bethe's ansatz appeared to be  escaped from physicists' attention that time.
It was only 30 years later that in 1963 Lieb and Liniger\cite{2}
first   solved the one-dimensional (1D)  many-body problem of
delta-function interacting bosons by Bethe's hypothesis. The exact
solution for the delta-function interacting Bose gas was given in
terms of the wave numbers $k_i$ ($i=1,\ldots, N$) satisfying a set
of Bethe ansatz equations, called the Lieb--Liniger equations.
The spectrum was given by summing up all $k_i^2$.
In fact, the Bethe ansatz equations describe the roles of individual
particles in many-body corrections. The significance of quantum
integrability is such that the roles of individual particles enable
one to precisely access full aspects of many-body physics for a
particular university class of many-body systems, for example, spin
chains, interacting quantum gases, strongly correlated electronic
systems, Kondo impurity problems, Gaudin magnets, etc.
Once concerning ensemble statistics, we have to  properly distinguish physical origin of distinguishable and indistinguishable particles.
At room temperature, the molecules in the air can be treated as
billard balls   that occasionally collide with each other.
The size of those molecules is much smaller than the mean distance between them.
The particles are distinguishable.
However, according to de Broglie matter wave theory, the thermal wavelength of a moving particle is given by a simple formula: $\lambda_{\rm dB} =\sqrt{2\pi\hbar^2/ (mk_{\rm B}T)}$, where $m$ is the  mass of the particle, $\hbar$ is the Plank constant, $k_{\rm B}$ is the Boltzmann  constant, and $T$ is the temperature.
The temperature becomes very low, the thermal wave length increases,
and  the wave packets start to overlap. Thus, the particles are
indistinguishable below  the degenerate temperature. The quantum
statistics play an essential role under such a  degenerate
temperature.
Below the degenerate temperature, there are thus fundamental differences
between the properties of fermions (with spins of $1/2$, $3/2$, $\ldots$ ) and bosons (with spins of $0$, $1$, $2$, $\ldots$).
All fermions must obey the Pauli exclusion principle, which means that they cannot occupy the same quantum state.
However, as bosons are not subject to the same restrictions, they
can collapse under suitable conditions into the same quantum ground
state, known as a Bose--Einstein condensate.
The Lieb--Linger model provides an ideal ensemble to understand the physics resulting from the distinguishable and indistinguishable nature of classical and quantum particles.

Towards a deeper understanding  of the physics of the Lieb--Liniger
gas,  a  significant next step was made by Yang C N and Yang C P on the
thermodynamics of this many-body problem  in 1969.\cite{3}
They for the first time present a grand canonical description of
the model in equilibrium. In fact, there are many microscopic states
for an equilibrium state of the system at finite temperatures.
The minimization of Gibbs free energy gives rise to the so-called Yang--Yang equation that determines
the true physical state in an analytical way.
In the grand canonical ensemble, the total number of particles can be associated with chemical potential $\mu$.
The temperature is associated with the entropy $S$, counting thermal disorder.
This canonical Yang--Yang approach marks a significant step to the exact solutions of finite temperature many-body physics.
It shows the subtlety of vacuum fluctuation, interaction effect, excitation modes, criticality, quantum statistics, thermalization, dynamics, correlations, and  Luttinger liquid.

The name Bethe's hypothesis was coined by Yang C N and Yang C P in
the study of the Heisenberg spin chain.
The Bethe ansatz is now well accepted as a synonym of quantum integrability.
>From solving eigenvalue problem of spin-1/2 delta-function
interacting Fermi gas, Yang\cite{4} found that the many-body
scattering matrix can be reduced to a product of many two-body
scattering matrices, i.e., a necessary condition for solvability of
the 1D many-body systems.
The two-body scattering matrix satisfies a certain intertwined relation, called Yang--Baxter equation.
This seminal work has inspired a great deal of developments in physics and mathematics.
The Yang--Baxter relation was independently shown by Baxter as the
conditions for commuting transfer matrices in two-dimensional
statistical mechanics.\cite{5,6}
For such exactly solved models, the energy eigen-spectrum of the
model Hamiltonian can be obtained exactly in terms of the Bethe
ansatz equations, from which physical properties can be derived via
mathematical analysis.
The Lieb--Liniger Bose gas\cite{2} and Yang--Gaudin
model\cite{4,7} are both notable Bethe ansatz integrable models.

Yang--Baxter solvable models have flourished into majority in physics since last 70's.
Later it turned out that the Yang--Baxter integrability plays an
important role in physical and mathematical
studies.\cite{8,9,10,11,12,13,14,15}
The Bethe ansatz approach has also found success in the realm of condensed matter physics, such as Kondo impurity problems, BCS pairing models, strongly correlated electron systems and spin ladders, cold atoms, quantum optics, quantum statistical mechanics, etc.
The Yang--Baxter equation has led to significant developments in mathematics, such as 2D conformal field theory, quantum croups, knot theory, 2D statistical problems, lattice loop models, random walks, etc.
Recent research showed that there exists a remarkable connection between conformal field theory and Yang--Baxter integrability of 2D lattice models.
Remarkably, conformal field theory has led to the theory of vertex operator algebras, modular tensor categories, and algebraic topology in connection to new states of matter with topological order, such as fractional quantum Hall effect, topological insulators, etc.
In this elementary introduction to the exactly solvable Lieb--Liniger model, we will discuss the rigorousness of mathematical integrability and  the novelty of quantum many-body effects that comprise the beautiful cold world of many-body systems  in the lab.
The content of this article involves understanding the fundamental many-body physics  through the  model of Lieb--Liniger Bose gas.

The paper is organized as follows.
In Section~2, we present a rigorous derivation of the Bethe ansatz
for the Lieb--Liniger Bose gas.
In this section we show how a field theory problem reduces to a quantum-mechanical many-body systems following the Lieb--Liniger's derivation.
In Section~3, the properties of the ground state are discussed,
including ground-state energy, excitations, cooperative and
collective features, Luttinger parameter, etc.
In Section~4, we introduce the Yang--Yang grand canonical approach
to the finite temperature physics of the Lieb--Linger gas.
In this section, we demonstrate how the Yang--Yang equation encode the subtle Bose--Einstein statistics, Fermi--Dirac statistics, and Boltzmann statistics.
In particular, we give an insightful understanding of quantum criticality.
In Section~5, we briefly review  some of  recent  experimental
measurements related to the Lieb--Liniger Bose gas from which one
can conceive the beauty of the integrability.

\section{Bethe ansatz for the Lieb--Liniger Bose gas}

\subsection{Wavefunction}

We start with introduction of canonical quantum Bose fields
$\hat\psi(x)$   satisfying the following  commutation relations:
\begin{eqnarray*}
\begin{split}
 &\big[\hat\psi(x),\hat\psi^\dag(y)\big]=\delta(x-y),\\
 &\big[\hat\psi(x),\hat\psi(y)\big]=
 \big[\hat\psi^\dag(x),\hat\psi^\dag(y)\big]=0.
\end{split}
\end{eqnarray*}
The Hamiltonian of the 1D single component bosonic quantum gas  of
$N$ particles in a 1D box with length $L$ is given by\cite{2}
\begin{eqnarray}
  \el{1}
  \hat{H} =
  \frac{\hbar^2}{2 m}\int_0^L {\rm d} x~\partial_x\hat{\psi}^\dag
  \partial_x \hat\psi
  + \frac{g_{\rm 1D}}{2} \int_0^L {\rm d} x ~\hat\psi^\dag
  \hat\psi^\dag\hat\psi\hat\psi,
\end{eqnarray}
where $m$ is the mass of the bosons, $g_{\rm 1D}$ is the coupling
constant which is determined by the 1D scattering length $g_{\rm
1D} = -2 \hbar^2/m a_{\rm 1D}$. The scattering length is given by
$a_{\rm 1D}=\left( -a_{\perp }^{2}/2a_{\rm s}\right) \left[
1-C\left(a_{\rm s}/a_{\perp }\right)
\right]$.\cite{16,16b,17}
Here the numerical constant $C\approx 1.4603$.
The model~(\eref{1}) presents the second-quantized form of the
Lieb--Liniger Bose gas with contact interaction.\cite{2}

In order to process Lieb and Liniger's solution, we first define the
vacuum state in the Fock space  as $\hat\psi(x)|0\rangle =0,
~x\in\mathbb{R}$ with $\langle 0\mid 0\rangle=1$.
The equation of motion for the field $\hat \psi(x)$ is given by the Heisenberg equation ${\mathrm i} \partial_t \hat \psi(x)=[\hat H, \hat\psi(x)]$.
It follows that the corresponding equation of the motion for this model reads
\begin{eqnarray}\el{2}
 {\mathrm i} \partial_t \hat \psi(x)=-\partial_x^2\hat \psi(x)+2c \hat \psi^\dag(x) \hat\psi(x) \hat\psi(x).
\end{eqnarray}
Considering $\hat \psi(x)$ as a classical field, this equation of
motion reduces to a non-linear Schr\"odinger equation of the
classical field theory.
Moreover, it is easy to show that the particle number operator $\hat N$ and the momentum operator  $\hat P$
\begin{eqnarray}\el{3}
 \hat N=\int_0^L \hat \psi^\dag \psi {\rm d}x,~~~\hat P=-\frac{\mathrm i}2 \int_0^L \bigg\{
  \big[\partial_x,\hat \psi^\dag(x)\big]
  \hat\psi(x) \bigg\}{\rm d}x~~~
\end{eqnarray}
are  commutative with the Hamiltonian~(\eref{1}), i.e., $[\hat H,
\hat N]=0$ and $[\hat H, \hat P]=0$. They are among the conserved
quantities of this model.
The eigenfunction of the $N$-particle state $|\varPsi\rangle $ for
the operators $  \hat{H} $, $  \hat{N} $, and $  \hat{P} $ is given
by
\begin{eqnarray}\el{4}
 |\varPsi\rangle
 =\frac 1{\sqrt{N!}} \int_0^L {\rm d}^N \bm x~
 \varPsi(\bm x)|\bm x\rangle ,
\end{eqnarray}
where $\bm x=\{x_1,x_2, \ldots, x_N\}$ and $$|{\bm x}\rangle
=\hat\psi^\dag(x_1) \hat\psi^\dag(x_2) \cdots \hat\psi^\dag(x_N)
|0\rangle .$$
Here $x_j$ is the coordinate position of the $j$-th particle.
For the bosons,  the first quantized wavefunction $\varPsi$ is
symmetric with respect to any exchange of  two particles in space
$\bm x$, namely,
\begin{eqnarray}\el{5}
 \varPsi(\ldots, x_\xi,\ldots, x_\eta,\ldots)
 =\varPsi(\ldots, x_\eta,\ldots,x_\xi,\ldots).
\end{eqnarray}

In the following discussion, we set $\hbar = 2 m = 1$ and $c = mg_{\rm 1D}/\hbar^2$.
After some algebra, one can  find that the eigenvalue problem of the
Schr\"odinger equation $\hat{H} |\varPsi\rangle =E   |\varPsi\rangle
$ in $N$-particle sector reduces to the  quantum-mechanical
many-body problem which is described by the Schr\"odinger equation
${H}\varPsi(\bm x)=E\varPsi(\bm x)$ with the first quantized form of
the Hamiltonian
\begin{eqnarray}\el{6}
  H = - \sum_{i = 1}^N \frac{\partial^2}{\partial x_i^2} + 2 c \sum_{i < j}^N
  \delta ( x_i - x_j).
\end{eqnarray}
This many-body Hamiltonian describes $N$ bosons with $\delta$-function interaction in one dimension, called the Lieb--Liniger model.
This is a physical realistic model in quantum degenerate gases with
s-wave scattering potential.
In the dilute quantum gases, when the average distance between
particles is much larger than the scattering length, the s-wave
scattering between two particles at $x_\xi$ and $x_\eta$ have the
following short distance behavior:\cite{18}
\begin{eqnarray}\el{7}
  &\textstyle
  \varPsi'(0^+)-\varPsi'(0^-)=-\frac1{a_{\rm 1D}}[\varPsi(0^+)+\varPsi(0^-)],
\end{eqnarray}
where $\varPsi(x)$ is the relative wavefunction of
the two particles and $x$ is the relative distance between the two
particles, i.e.,  $x=x_\eta-x_\xi$. In the above equation, the prime
denotes the derivative with respect to $x$.

In the model (\eref{6}), the interaction only occurs when two
particles contact with each other.
Following the Bethe ansatz,\cite{1} we can divide the wavefunction
into $N!$ domains according to the positions of the particles
$\varTheta({\cal Q}): x_{{\cal Q}_1}< x_{{\cal Q}_2}<\cdots<
x_{{\cal Q}_N}$, where $\cal Q$ is the permutation of number set
$\{1,2,\ldots, N\}$.
The wavefunction can be written as $\varPsi(\boldsymbol{x})=\sum_{\cal Q}\varTheta({\cal Q}) \psi_{\cal Q}(\boldsymbol{x})$.
Considering the symmetry of bosonic statistics, all the $\psi$ in
different domain  ${\cal Q}$ should be the same, i.e., $\psi_{\cal
Q}=\psi_{\textbf{1}}$, where we denote the unitary element of the
permutation group as $\textbf{1}=\{1,2,\ldots,N\}$.
Lieb and Linger wrote the wavefunction for the model (\eref{6}) as
the superposition of $N!$ plane waves\cite{2}
\begin{eqnarray}\el{8}
 \label{ew1}
 \psi_{\textbf{1}} =\sum_{\cal P}A({\cal P}){\rm e}^{{\mathrm i}(k_{{\cal P}_1}x_{1}+\cdots+k_{{\cal P}_N}x_{N})},
\end{eqnarray}
where $k$'s are the pseudo-momenta carried by the particles under a
periodic boundary condition.

Indeed, after solving Schr\"odinger equation ${H} \varPsi(\bm
x)=E\varPsi(\bm x)$, we can obtain the same s-wave scattering
boundary condition (\eref{7}) that provides the two-body scattering
relation among the coefficients $A({\cal P})$
\begin{eqnarray}\el{9}
 A({\cal P}')=\frac{k_{{\cal P}_j}-k_{{\cal P}_{j+1}}+{\mathrm i}c}{k_{{\cal P}_j}-k_{{\cal P}_{j+1}}-{\mathrm i}c} A({\cal P}),
\end{eqnarray}
where ${\cal P}'$ is a permutation obtained by exchanging ${\cal P}_j$ and ${\cal P}_{j+1}$, i.e., ${\cal P}'=\{{\cal P}_1, \ldots, {\cal P}_{j-1}, {\cal P}_{j+1}, {\cal P}_j, {\cal P}_{j+2},$ $\ldots, {\cal P}_N\}$.
The scattering process in Eq.~(\eref{9}) implies that any two-body
scattering with momenta $k_{\rm a}$ and $k_{\rm b}$ leads to an
anti-symmetric phase shift $A({\cal P}')={\rm e}^{-{\mathrm
i}\theta(k_{{\cal P}_j}-k_{{\cal P}_{j+1}})} A({\cal P})$,
\begin{eqnarray}\el{10}
 \theta(k_{\rm a}-k_{\rm b})=2{\rm arctan}\bigg(\frac{k_{\rm a}-k_{\rm b}}{c}\bigg).
\end{eqnarray}
When $c\neq0$, all the quasi-momenta are different.
If there are two equal pseudo momenta $k_{\rm a}=k_{\rm b}$, we can
prove that the wavefunction $\psi_{\textbf{1}}=0$.
In general, equation~(\eref{9}) gives the two-body scattering matrix
$\hat S$, $A({\cal P}')=\hat S A({\cal P})$ for a quantum many-body
system. Yang\cite{4} proved that, if the two-body scattering matrix
$\hat S$ satisfies the following equation:
\begin{eqnarray}\el{11}
\hat S_{12}(\lambda-\mu) \hat S_{13}(\lambda) \hat S_{23}(\mu) =
\hat S_{23}(\mu) \hat S_{13}(\lambda) \hat S_{12}(\lambda-\mu),
\end{eqnarray}
then the system is integrable.
This relation was independently found by
Baxter\cite{5,6} in studying two-dimensional
statistical models.
Nowadays, equation (\eref{11}) is called Yang--Baxter equation. The
Yang--Baxter equation guarantees that the multi-body scattering
process can be factorized as the product of many two-body scattering
processes. This factorization reveals the nature of the
integrability, i.e., no diffraction in outgoing waves.
For the Lieb--Linger model (\eref{6}), $\hat S$ matrix is a scalar
function so  that the scattering matrix satisfies the Yang--Baxter
equation trivially.
In the scattering process from $\textbf{1}$ to ${\cal P}$, the
multi-body scattering matrix is defined by $A({\cal P})=\hat S({\cal
P}\boldsymbol{k})A(\textbf{1})$.
By using Eq.~(\eref{9}), the multi-body scattering matrix of this
model is given by
\begin{eqnarray}\el{12}
 \hat S({\cal P}\boldsymbol{k})=\prod_{{\cal P}_j<{\cal P}_l}
 \frac{k_{{\cal P}_j}-k_{{\cal P}_l}-{\mathrm i}c}{k_{{\cal P}_j}-k_{{\cal P}_l}+{\mathrm i}c}.
\end{eqnarray}
With the help of Eq.~(\eref{12}), the eigen wavefunction of the
system is given by
\begin{eqnarray}\el{13}
 \varPsi(\boldsymbol{x})=\sum_{\cal Q,{\cal P}} \varTheta({\cal Q}) \bigg(\prod_{{\cal P}_j<{\cal P}_l}
 \frac{k_{{\cal P}_j}-k_{{\cal P}_l}-{\mathrm i}c}{k_{{\cal P}_j}-k_{{\cal P}_l}+{\mathrm i}c}\bigg){\rm e}^{{\mathrm i}{\cal Q}\boldsymbol{x} \cdot {\cal P}\boldsymbol{k}}.
\end{eqnarray}

\subsection{Bethe ansatz equations}

Submitting  the periodic  boundary conditions
$\varPsi(\ldots,x_\xi=0,\ldots) ={\rm e}^{{\mathrm i}\alpha}
\varPsi(\ldots,x_\xi=L,\ldots)$ into the wavefunction (\eref{13}),
we can find that the pseudo momenta $k_l$ satisfies the following
Bethe ansatz equations (BAE):
\begin{eqnarray}\el{14}
 {\rm e}^{{\mathrm i} k_i L} = -{\rm e}^{-{\mathrm i}\alpha} \prod_{j = 1}^N \frac{k_i - k_j + {\mathrm i} c}{k_i - k_j - {\mathrm i} c},~~~i=1,2,\ldots,N,
\end{eqnarray}
which  are  called  the Lieb--Liniger equations.
When $\alpha=\pi$, the wavefunction is anti-periodic; while when $\alpha=0$, it is periodic.
In the following discussion, we only consider the periodic boundary conditions.

Since partial number $\hat N$ and momentum $\hat P$ are conversed quantities of the Lieb--Liniger model, the Hamiltonian together with $\hat N$ and $\hat P$ can be simultaneously diagonalized.
For the eigenstate (\eref{13}), the corresponding particle number
$\langle {\hat N}\rangle=N$. For a given set of quasi-momenta
$\left\{ k_j\right\}$, the total momentum and the energy of the
system are obtained as
\begin{eqnarray}\el{15}
 P=\langle \hat P\rangle=\sum_{j}^N k_j,~~~
 E=\langle \hat H\rangle=\sum_{j}^N k^2_j.
\end{eqnarray}

The solutions to the  BAE~(\eref{14}) provide complete spectra of
the Lieb--Linger model.
The physical solutions to the Bethe ansatz equations require that all the pseudo momenta are distinct to each other.
The BAE (\eref{14}) can be written in the form of phase shift
function $\theta(k)$ as
\begin{eqnarray}\el{16}
  2 \pi \frac{I_i}L = k_i  + \frac1L \sum_{j = 1}^N \theta \bigg( \frac{k_i - k_j}{c}
  \bigg), \label{lbae}
\end{eqnarray}
where $\{I_i\}$ are the quantum numbers of pseudo momenta.
If $N$ is odd, these quantum numbers  are integers, whereas they are half odd integers  when $N$ is even.
For a given set of  quantum numbers $\{I_i\}$, there is a unique set of real values $\{ k_i \}$ for  $c>0$.
These quantum numbers are independent of coupling constant $c$.
The total momentum can be expressed as
\begin{eqnarray}\el{17}
  P = \frac{2\pi}{L}\sum_{i = 1}^N I_i=2\ell \pi/L,\,\, \ell =0,\pm 1,\,\pm2,\,\ldots
\end{eqnarray}
For the ground state, $P=0$, where all quasi-momenta $\left\{ k_i\right\}$ are located in an interval $\left(-Q, Q \right)$.
Here $Q$ is the cut-off.
For the ground state, the quantum numbers are given by
\begin{eqnarray}\el{18}
I_j=-\frac{N-1}{2}+j-1,\,\,\,j=1,\ldots, N.
\end{eqnarray}

\section{Ground-state energy, excitations, and correlations}

In the thermodynamic limit, i.e., $N, L \to \infty$ and the particle
density $n=N/L $ is a constant, the BAE (\eref{16}) can be written
in the integral form
\begin{eqnarray}\el{19}
 \rho(k)= \frac{1}{2\pi}+ \int_{-Q}^Q  a(k-q)\rho ( q) {\rm d} q,
 ~~~|k|<Q,
\end{eqnarray}
where $\rho(k)$ is the  density distribution function  of the quasi-momenta defined by the particle numbers in a small interval of $(k,k+\Delta k)$, i.e.,
\begin{eqnarray}
  \rho(k) =\lim_{\Delta k\to0}\frac{1}{L\Delta k}.
  \nonumber
\end{eqnarray}
Here the  cut-off  $Q$ is the ``Fermi point"  of the pseudo momenta.
It is  determined by the particle density $n = \int_{-Q}^Q \rho ( k) {\rm d} k$.
Energy per length can also be written as
\begin{eqnarray}\el{20}
\frac{E}{L}= \int_{-Q}^Q \rho (k) k^2 {\rm d} k.
\end{eqnarray}

For the ground state, the energy depends on the  dimensionless scale $\gamma=Lc/N$.
Let us make a scaling transformation
\begin{eqnarray}\el{21}
k=Qx,\,\, c=Q\lambda,\,\, \rho(Qx) =g(x),
\end{eqnarray}
we find the ground-state energy  per particle,
\begin{eqnarray}\el{22}
\frac{E}{N}=\frac{\hbar^2 n^2}{2m}e_0(\gamma) ,
\end{eqnarray}
with $e_0(\gamma) =
\frac{\gamma^3}{\lambda^3}\int_{-1}^{1}x^2g_0(x)\zd x$.
Here the distribution function $g_0(x)$ is determined  by
\begin{eqnarray}\el{23}
 g_0(x)=\frac{1}{2\pi} +\frac{\lambda}{\pi}
 \int_{-1}^1 \frac{g_0(y)}{\lambda^2+(x-y)^2} {\rm d}y,
\end{eqnarray}
with the cut-off condition $\gamma\int_{-1}^1g_0(x) \zd x=\lambda$.
Equation (\eref{23})  is the standard inhomogeneous Fredholm
equation, which is well understood in mathematics.
The Fredholm equation can be numerically solved.
We will further study the ground-state energy later.

\subsection{Ground state: from cooperative to collective}

For the ground state, the competition of kinetic energy and interaction energy is represented by the dimensionless coupling strength $\gamma$.
When $\gamma=0$, the system is free bosons, and all the particles are condensed at the zero momentum state; while if there is a small coupling constant, all the $k$'s are distinct.
At the limit $\gamma=\infty$, the strong repulsion makes the quasi-momentum distribution the same as the one of the free fermions.

\subsubsection{Weak coupling limit: semicircle law}

It is very insightful to examine the physics of the model in weak coupling limit.
In this case the Bethe ansatz roots comprise the semi-circle law.
Gaudin firstly noticed such a kind of distribution,\cite{19}
followed by several groups.\cite{20,21}
The Fredholm equation (\eref{23}) is known as the Love equation for
the problem of the circular disk condenser.
By using Hutson's method, Gaudin found the density distribution function and energy density
\begin{eqnarray}\el{24}
g_0(t)& \approx &\frac{Q}{2\pi c}\left(1-t^2 \right)^{1/2} \nonumber \\
&& +\,\frac{1}{4\pi^2} \left( 1-t^2\right)^{-{1}/{2}}\left( t\ln \frac{1-t}{1+t}+\ln \frac{16\pi e Q}{c} \right),\nonumber \\
 e& =& n^3\bigg(\gamma - \frac{4}{3 \pi} \gamma^{3 / 2}\bigg).
\end{eqnarray}
This result coincides with the perturbative calculation by using the Bogoliubov method.

This problem can also be solved from the original BAE (\eref{14}).
In the weak coupling limit, i.e., $Lc\ll 1$, the quasi-momenta $k_j$ are proportional to the square root of $c$ and $c/(k_j-k_l)$ is a small value.
Up to the second order of $c$, the BAE (\eref{14}) is expanded
as\cite{21}
\begin{eqnarray}\el{25}
 q_j=\sum_{l\neq j}^N\frac{1}{q_j-q_l},
\end{eqnarray}
where $q_j=k_j\sqrt{L/2c}$.
If we define a function ${H}_N(q)=\prod_{i=1}^N(q-q_j)$, we can find
that $F(q_j)=0$ for the polynomial $F(q)\equiv {H}''_N(q)-2q {H}'_N(q)$.
Both $F(q)$ and ${H}_N(q)$ are the polynomial of degree $N$ and
$F(q_j)={H}_N(q_j)=0$, and $F(q)$ and ${H}_N(q)$ are proportional to
each other.
At the large order of $q$, $H(q)=q^N+\cdots$ and
$F(q)=-2Nq^N+\cdots$, so that we have $F(q)=2N{H}_N(q)$. It follows
that
\begin{eqnarray}\el{26}
 {H}''_N(q)-2q{H}'_N(q)+2N{H}_N(q)=0.
\end{eqnarray}
The solution of this differential equation is a Hermite polynomial
and $q_j$ is the root of the corresponding polynomial of degree $N$,
i.e., ${H}_N(q)=0$.
We set the order of $q_j$ as $q_j<q_{j+1}$, then we have
$(2N+1-q_j^2)^{1/2}>\pi/(q_{j+1}-q_j)>(2N+1-q_{j+1}^2)^{1/2}$.
The distribution function,
$\rho(k)=\lim_{L,N\to\infty}\frac1{L(k_{j+1}-k_j)}$, is thus
determined by\cite{22}
\begin{eqnarray}\el{27}
 \rho(k) \approx \frac{1}{\pi}\sqrt{\frac n{c}}\bigg(1-\frac{k^2}{Q^2}\bigg)^{1/2}+O\bigg(\frac{1}{Lnc}\bigg),
\end{eqnarray}
where the cut-off $Q=2\sqrt{nc}$ is obtained by $\int_{-Q}^Q \rho(k){\rm d}k=n$.
We can see that the quasi-momentum distribution function satisfies
the semi-circle law where the cut-off is the radius of this circle
(see Fig.~\fref{f1}{(a)}).
The leading order of energy (\eref{24}) was also found based on the
pseudo momentum distribution (\eref{27}).


\begin{center}
\fl{f1}\includegraphics[width=\linewidth]{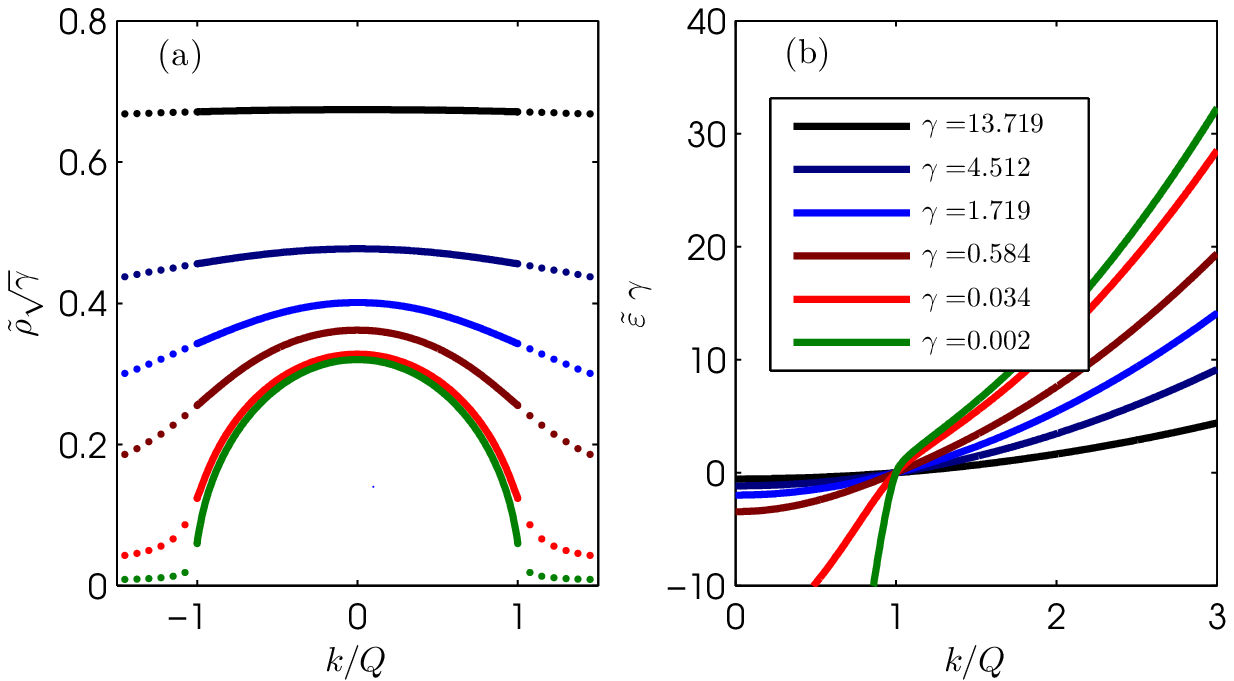}\\[5pt]
\parbox[c]{8.0cm}{\footnotesize{\bf Fig.~1.}
 Densities and dressed energies of pseudo momenta for the ground state.
(a) Solid lines: the dimensionless densities of the pseudo momenta,
$\tilde \rho(k)=\rho(k)/c$ obtained from Eq.~(\eref{23}); dotted
lines: the corresponding dimensionless hole densities, $\tilde
\rho_{\rm h}(k)=\rho_{\rm h}(k)/c$. When the coupling strength is
small, the distribution function $\tilde \rho$ meets a semi-circle
law (\eref{27}). For the strong coupling limit, i.e., $\gamma\gg 1$,
the distribution function gradually becomes flatter and flatter, and
approaches $\rho(k)\approx 1/2\pi$.
 (b) Dimensionless dressed energy is defined by $\tilde
\varepsilon(k)= \varepsilon(k)/c^3$, which is obtained from the
dressed energy equation~(\eref{64}).}
\end{center}
\vspace*{2mm}

\subsubsection{Strong coupling limit: fermionzation}

In the strong repulsion limit, the gas is known as the Tonks--Girardeau (TG) gas.
In realistic experiment with cold atoms, it is practicable to
observe the quantum degenerate gas with the strong coupling
regime.\cite{23,24}
In the  regime $\gamma \to \infty$, the Bose--Fermi mapping
method\cite{18} can map out  the ground-state properties of the
Bose gas through the wavefunction of the non-interacting fermions.
For  finitely strong interaction, $\gamma \gg 1$, we can obtain the
ground-state pseudo momenta from the Bethe asnatz
equations~(\eref{14})\cite{25,26}
\begin{eqnarray}\el{28}
k_j
 &=&2\pi \frac{I_j}L\bigg(1-\frac2\gamma+\frac4{\gamma^2}-\frac{8}{\gamma^3}\bigg)
 \nonumber\\
&&+\,\frac{4\pi^3}{3c^3L^4}\bigg[\bigg(N+\frac12-j\bigg)^4-\bigg(\frac12-j\bigg)^4\bigg]
+O\bigg(\frac1{c^5}\bigg),~~~~~~~
\end{eqnarray}
where $I_j$ take the ground-state quantum numbers
$I_j=-\frac{N-1}{2}+j-1,\,\,\,j=1,\ldots, N$.
With the help of Eq.~(\eref{28}), the ground-state energy of the
strong repulsive gas is given by
\begin{eqnarray}\el{29}
 \frac{E}{L}\approx\frac{\pi^2}{3} n^3\bigg[ 1 - \frac{4}{\gamma} +
  \frac{12}{\gamma^2} + \frac{32}{\gamma^3} \bigg( \frac{\pi^2}{15}-1
  \bigg) \bigg)\bigg].
\end{eqnarray}
This asymptotic result fits well with the numerical result obtained
by solving the integral BAE~(\eref{19}) (see Fig.~\fref{f2}{(a)}).
The ground-state energy (\eref{29}) can also be obtained from the
integral BAE (\eref{19}) by strong coupling expansion
method.\cite{25,27}
We see that for $\gamma\to\infty$ the leading order of the ground-state energy reduces to that of the free fermions $e_{\rm f}=\pi^2 n^3/3$.
One can also calculate other important properties such as compressibility, sound velocity, and Luttinger parameter, once we know the dimensionless function
$e_0(\gamma)$.

Very recently, Ristivojevic obtained high-precision ground-state
distribution function for the strong coupling regime\cite{28}
\begin{eqnarray}\el{30}
 \rho(k)& =&\frac{1}{2 \pi }+\frac{1}{\pi ^2 \lambda }+\frac{2}{\pi ^3 \lambda ^2}+\frac{12-\pi ^2}{3 \pi ^4 \lambda ^3}+\frac{8-2\pi ^2}{\pi ^5 \lambda ^4}
 \nonumber\\
 &&-\,\frac{k^2}{Q^2}\bigg(\frac{1}{\pi ^2 \lambda ^3}+\frac{1}{\pi ^3 \lambda ^4}\bigg)+{\cal O}(\lambda^{-5}).
\end{eqnarray}
The ground-state energy per unit length and particle density are given by
\begin{eqnarray}\el{31}
 \frac{E}{L} &=&\frac{\lambda ^3}{3 \pi  c^3}
\bigg(1+\frac{2}{\pi  \lambda }+\frac{4}{\pi ^2 \lambda ^2}
+\frac{120-28\pi ^2}{15\pi ^3 \lambda ^3}\notag\\
&&+\,\frac{80-26\pi ^2}{5\pi ^4 \lambda ^4}\bigg),\\
\el{32} n &=&\frac{\lambda }{c \pi }\bigg(1+\frac{2}{\pi  \lambda
}+\frac{4}{\pi ^2 \lambda ^2}+\frac{24-4\pi ^2}{3\pi ^3 \lambda
^3}\notag\\
&&+\,\frac{48-14\pi ^2}{3\pi ^4 \lambda ^4}\bigg),
\end{eqnarray}
where $\lambda = (\gamma+2)/\pi -4\pi/(3 \gamma ^2)+16\pi/(3 \gamma ^3)$.
The result  $e_0=E/(Ln^3)$ is plotted in Fig.~\fref{f2}.

\vspace*{2mm}

\begin{center}
\fl{f2}\includegraphics[width=\linewidth]{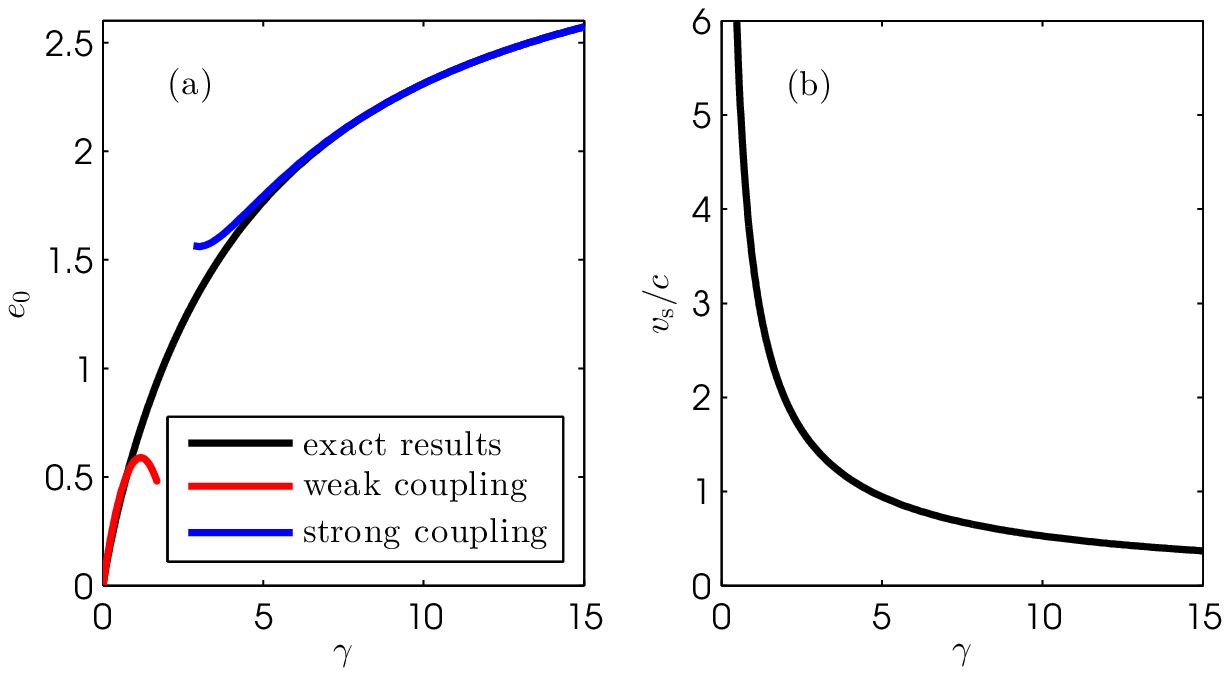}\\[5pt]
\parbox[c]{8.0cm}{\footnotesize{\bf Fig.~2.}
The ground state energy and sound velocity for different coupling
strength $\gamma$.
 (a) The ground energy density is a monotonic increasing function of $\gamma$ when the linear density is fixed.
 Black line: the exact numerical result from Eq.~(\eref{23}).
 Red line: the weakly coupling expansion result~(\eref{24}).
 Blue line: the result obtained by the strongly coupling expansion~(\eref{31}).
 (b) The sound velocity is a monotonic decreasing function of $\gamma$.
  In the weakly coupling limit, the ratio turns to be infinite, while in the strong coupling limit, $v_{\rm s}/c \to=2\pi/\gamma$ (see
  Eq.~(\eref{45})).}
  \end{center}

\subsubsection{Elementary excitations: collective motion}
As discussed in previous section, the eigenstates of the model are described by the quantum numbers $\{I_j\}$.
We define $I_j$ as a function of the psudomomenta, i.e.,  $I(k_j)=I_j$, thus we have
\begin{eqnarray}\el{33}
 \frac{I(k)}{L}=\frac{k}{2\pi} +\frac1L \sum_{j=1}^N \frac{\theta(k-k_j)}{2\pi}.
\end{eqnarray}
Usually we define the occupied Bethe ansatz roots as particles,
while the unoccupied roots as holes.
For a hole quasimomentum $k_{\rm h}$, the quantum number $I_{\rm
h}=I(k_{\rm h})\in \mathbb{Z}+(N+1)/2$. We demonstrate the quantum
numbers $I_j$ for the ground state and excited states in
Figs.~\fref{f3}{(a)}--\fref{f3}{(d)}.
%
For the ground state, there is no hole below the quasi-Fermi
momentum, i.e.,  $|k|<Q$, whereas the quantum numbers  $I$'s for the
holes site outside the interval $[I_1,I_N]$ (see
Fig.~\fref{f3}{(a)}).

In the thermodynamic limit, the pseudo momentum distribution
function $\rho_{0}(k)$ for the ground state  is determined by Eq.
(\eref{19}) with the cut-off of $Q$.
The ground-state energy and the particle density can be regarded as the function of the cut-off $Q$, i.e., $e(Q)$ and $n(Q)$.
The changes over the configuration of quantum numbers for the ground state give rise to excited states.
Figure~\fref{f3}{(d)} shows such an excitation where the particle
with a quasimomentum  $k_{\rm h}$ below the pseudo Fermi point is
excited outside the Fermi point with a new quasimomentum $k_{\rm
e}$.

\vspace*{2mm}

\begin{center}
\fl{f3}\includegraphics[width=\linewidth]{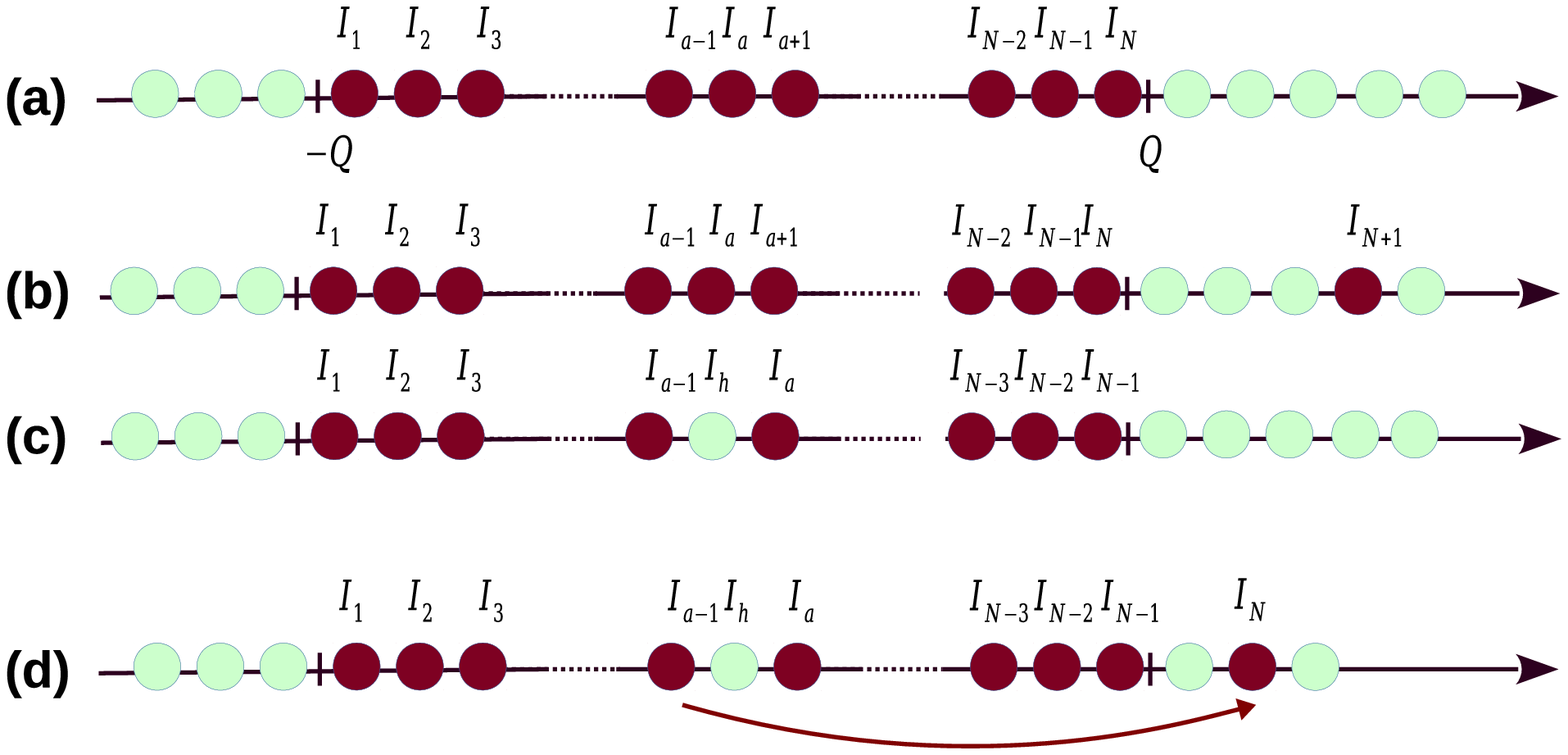}\\[5pt]
\parbox[c]{8.7cm}{\footnotesize{\bf Fig.~3.}
Schematic diagrams of the ground state and elementary excitations.
 (a) Quantum numbers for the ground state. The quantum numbers for the ground state are symmetric around the the origin. The largest quasimomentum denotes the ``Fermi points" $\pm Q$.
 (b) Configuration of adding a particle near the right Fermi point with the quantum number $I_{N+1}$, so that the total number of particles is $N+1$.
 (c) A hole excitation. The hole at $I_{\rm h}$ is created  so that the total number of particles is $N-1$.
In panels (b) and (c),  the parities  of  their quantum numbers are
changed from half-odd (or integers) to integer (or half-odds) due to
the changes of particle numbers.  (d) A single particle--hole
excitation. A particle at the position $I_{\rm h}$ is excited out of
the pseudo Fermi sea.   In this case, the total number of particles
is still $N$, and the parity of quantum numbers dose not change.}
\end{center}
\vspace*{2mm}

For a single-particle excitation, we decompose the pseudo momentum density into two parts, $\rho_t(k)=\bar\rho(k)+\frac1L \delta(k-k_{\rm e})$,
where the delta function term attributes to the excited particle and $\bar\rho(k)$ stands for the density below the cut-off pseudo momenta,  i.e., $|k|<Q$.
In the thermodynamic limit, $\bar\rho(k)$ satisfies the following integral equation:
\begin{eqnarray}\el{34}
\bar\rho(k)=\frac1{2\pi}+\int^{Q}_{-Q}a(k-k')
\bar\rho(k')\zd k' +\frac1L a(k-k_{\rm e}).
\end{eqnarray}
Here we choose a state which satisfies Eq.~(\eref{19}) with cut-off
$Q$ as a reference state, i.e.,
\begin{eqnarray}\el{35}
\rho_0(k)=\frac1{2\pi}+\int^{Q}_{-Q}a(k-k')\rho_0(k')\zd k'+\frac1L
a(k-k_{\rm e}).
\end{eqnarray}
The addition of the excited particle leads to a collective rearrangement of the distribution of the pseudo momenta of the $N-1$ particles.
We denote the difference of the pseudo momentum distribution functions between the excited and the reference state as $\Delta \rho(k)=\bar\rho(k)-\rho_0(k)$, where $\rho_0(k)$ is the pseudo momentum distribution of the reference state within $|k|<Q$.
By comparing the integral BAE (\eref{34}) and (\eref{19}), we can
find that $\Delta \rho$ satisfies the following equation:
\begin{eqnarray}\el{36}
 \Delta\rho(k)=\frac1L a(k-k_{\rm e})+\int^{Q}_{-Q}a(k-k') \Delta\rho(k').
\end{eqnarray}
The addition of the particle also leads to a shift of the cut-off over the psuado-Fermi point of the ground state of $N$-particle, i.e., $\Delta Q=Q-Q_{\rm G}$.
Here, $Q_{\rm G}$ is the cut-off for the ground state with the particle number $N=Ln_{\rm G}(Q_{\rm G})$, and $Q$ is that for the excited state with  $N=Ln_{\rm G}(Q)+L\int_{-Q}^Q {\rm d}k \Delta \rho(k)+1$.
In the thermodynamic limit, $\Delta Q$ is a small value. By comparing the formulas of particle numbers between the ground state and the excited state,  $\Delta Qn'_{\rm G}(Q_{\rm G})=-[\int_{-Q}^Q \Delta\rho(k){\rm d}k+1/L]$.
The prime denotes the derivative with respect to the the cut-off  $Q$.
The corresponding excited energy
\begin{eqnarray}\el{37}
 \Delta E(k_{\rm c})
 & = &L\int^{Q}_{-Q} \Delta\rho(k)k^2{\rm d}k+k_{\rm e}^2+ Le'(Q)\Delta Q
 \nonumber\\
 & = &L\int^{Q}_{-Q} \Delta\rho(k)(k^2-\mu){\rm d}k+(k_{\rm e}^2-\mu),
\end{eqnarray}
where $\mu$ is the chemical potential, and $\mu={\rm d} E/{\rm d}
N=e_{\rm G}'(Q_{\rm G})/n_{\rm G}'(Q_{\rm G})$.
Later we will further prove that the dressed energy can be expressed as
\begin{eqnarray}\el{38}
 \varepsilon(k)=k^2-\mu+\int^{Q}_{-Q}a(k-k') \varepsilon(k').
\end{eqnarray}
For convenience, we introduce a useful relation between the dressed energy and density.
Assuming that two functions $f(k)$ and $g(k)$ satisfy the following equations:
\begin{eqnarray}\el{39}
&&  f(k)=f_0(k)+\int_{-Q}^Q a(k-k') f(k') {\rm d}k',~~|k|< Q,\notag\\
&&  g(k)=g_0(k)+\int_{-Q}^Q a(k-k') g(k') {\rm d}k',~~|k|< Q,
\end{eqnarray}
where $f_0(k)$ and $g_0(k)$ are the driving terms of these integral equations.
Then we have the useful relation
\begin{eqnarray}\el{40}
  \int_{-Q}^Q f(k)g_0(k) {\rm d} k=\int_{-Q}^Q g(k)f_0(k) {\rm d} k.
\end{eqnarray}
By using Eq.~(\eref{40}), from Eqs.~(\eref{38}) and (\eref{39}), we
thus prove that the dressed energy is nothing but the excitation
energy of a particle,
\begin{eqnarray}\el{41}
& &\textstyle
 \varepsilon(k_{\rm e})=k_{\rm e} ^2-\mu+\int^{Q}_{-Q}a(k_{\rm e}-k') \varepsilon(k')
 \nonumber\\
 &\textstyle
 =&k_{\rm e}^2-\mu+L\int^{Q}_{-Q} (k^2_{\rm e}-\mu) \Delta\rho(k')=\Delta E(k_{\rm e}).
\end{eqnarray}

On the other hand, the hole excitations impose an additional condition, namely, the maximum momentum is $n\pi$.
Similarly,  one can prove that the excited energy reads
\begin{eqnarray}\el{42}
 \Delta E(k_{\rm h})=-\varepsilon(k_{\rm h}).
\end{eqnarray}
We plot the dressed energies for different coupling strengths in
Fig.~\fref{f1}{(b)}.
In order to see clearly excitation spectra, we need to calculate the
total momentum using Eq.~(\eref{17}).
For both the particle--hole excitation and the Type-II hole
excitation, the total particle numbers of particles do not change,
as shown in Figs.~\fref{f3}{(d)} and \fref{f4}{(b)}.
Therefore, the total momenta of the excited states are given  by
\begin{eqnarray}\el{43}
  P =n\pi -2\pi\int_{0}^{k_{\rm e,h}} \rho_0(k) {\rm d} k.
\end{eqnarray}

For the low-energy excitations, $k_e-Q$ is a very small value.
Therefore, the low-lying behavior can be described by a linear dispersion relation
\begin{eqnarray}\el{44}
 \Delta E=\varepsilon'_{\rm G}(Q)|k\pm Q|=v_{\rm s}|P|,~~~
 v_{\rm s}=\frac{\varepsilon'_{\rm G}(Q)}{2\pi \rho_{\rm G}(Q)},
\end{eqnarray}
where $v_{\rm s}$ is the sound velocity of the system.
In the above equation, $\pm$ corresponds to the excitations at left/right Fermi point, respectively.
Nevertheless, for $|k|>Q$,  the relation~(\eref{44}) is  the
dispersion relation for particle--hole excitations.
This linear spectra uniquely determine the universal Luttinger liquid behaviour at the low temperatures.
The essential feature resulted from the dispersion
relation~(\eref{44}) is such that the system exhibits the conformal
invariant in low energy sector.
We will further discuss this universal nature of the 1D many-body physics.
For the strong coupling, we can obtain the sound velocity from the ground-state energy through the relation $v_{\rm s}
  = \sqrt{\frac{L}{m n} \frac{\partial^2 E}{\partial L^2}}$, namely,
\begin{eqnarray}\el{45}
 v_{\rm s}\approx 2\pi n\bigg[1-\frac{4}{\gamma}+\frac{12}{\gamma^2}
 +\frac{16}{\gamma^3}\bigg(\frac{\pi^2}{3}-2\bigg)\bigg].
\end{eqnarray}
In Figs.~\fref{f2}{(a)} and \fref{f2}{(b)}, we present the
analytical and numerical results  for  the dimensionless energy
$\bar e(\gamma)=e/n^3$ and $v_s/c$, respectively.

\vspace*{2mm}

\begin{center}
\fl{f4}\includegraphics[width=\linewidth]{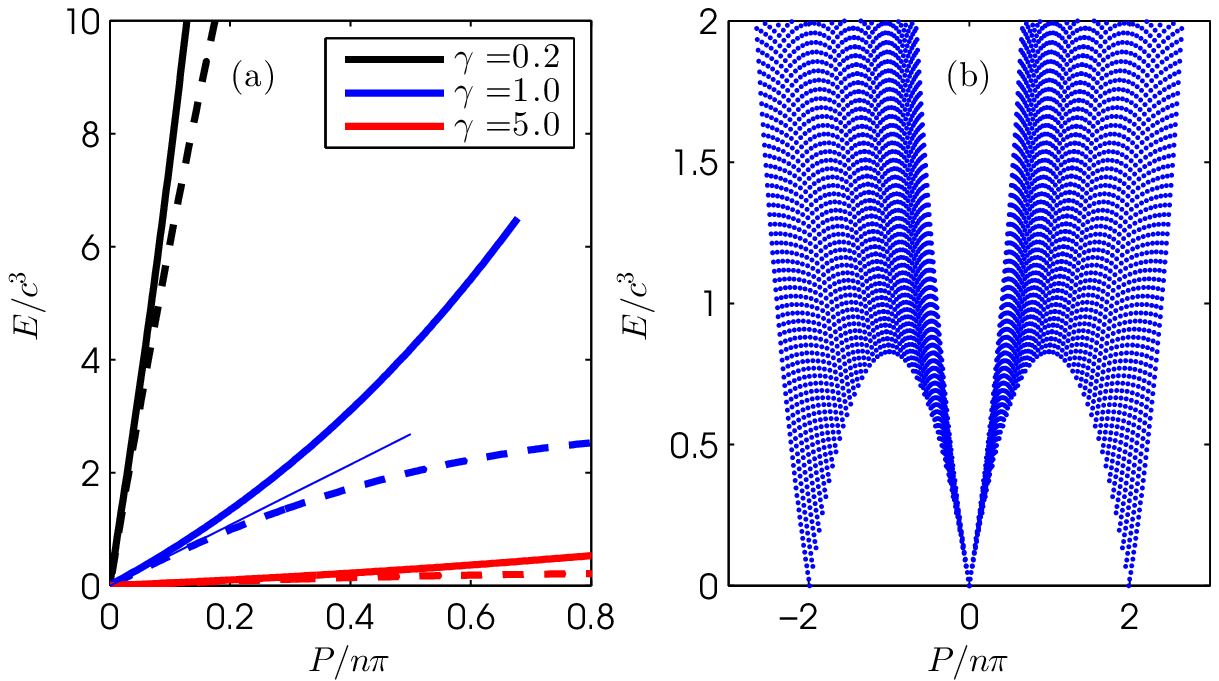}\\[5pt]
\parbox[c]{8.0cm}{\footnotesize{\bf Fig.~4.}
  Dispersion relations in elementary excitations:  adding one particle  and adding one hole excitation.  (a) Solid lines:  the spectra for adding one particle to the ground state corresponding to the configuration presented in Fig.~\fref{f3}{(b)}.
Dashed lines: the dispersion relations for the hole excitation, see
the configuration shown in Fig.~\fref{f3}{(c)}. For $\gamma=1$, the
thin blue line is calculated by using Eq.~(\eref{44}). We see that
both the particle and hole excitations comprise the dispersion
relations with the same velocity.    (b) The particle--hole
excitation  spectra, see Fig.~\fref{f3}{(d)}. The linear dispersion
relation is seen for the long-wavelength limit, i.e., the momentum
tends to be zero.}
\end{center}

\subsubsection{The super Tonks--Girardeau gas-like phase}

In regard of the strong interacting bosons in 1D, the super Tonks--Girardeau gas is particularly interesting.
It describes a gas-like phase of the attractive Bose gas which was
first proposed in a system of attractive hard rods by Astrakharchik
{\em et al.}\cite{29}
Batchelor and his coworkers showed its existence of such a novel
state in the Lieb--Liniger model with a strong attraction.\cite{30}
Due to the large kinetic energy inherited from the repulsive
Tonks--Girardeau gas, the hard-core behavior of the particles with
Fermi-like pressure prevents the collapse of the super TG phase
after the switch of interactions from repulsive to attractive
interactions.\cite{31,32,33,34}
In fact, the energy can be continuous in the limits $c\to \pm \infty$.
>From the Bethe ansatz equations (\eref{14}), near $c\to \pm \infty$,
the compressibility is given by
\begin{eqnarray}\el{46}
  \frac{1}{\kappa} = 2 \pi^2 n - \frac{16 \pi^2}{c} n^2 + \frac{80 \pi^2}{c^2}
  n^3 + \left( \frac{64}{3} \pi^2 - 320 \right)  \frac{n^4}{c^3}.
\end{eqnarray}
However, for the Lieb--Liniger gas with weak repulsive interaction ($0 < c \ll 1$),
compressibility is given by
\begin{eqnarray}\el{47}
  \frac{1}{\kappa} = 2 c - \frac{1}{\pi \sqrt{n}} c^{3 / 2}.
\end{eqnarray}
In contrast, the energy for the gas-like phase (super
Tonks--Girardeau gas) in  weak attractive interaction limit ($- 1
\ll c < 0$) is given by
\begin{eqnarray}\el{48}
  E = \frac{2 \pi^2}{3} n^2 - \left| c \right| n^2.
\end{eqnarray}
Thus, the compressibility of the super Tonks--Giraread gas with weak attraction is given by
\begin{eqnarray}\el{49}
  \frac{1}{\kappa} = 4 \pi^2 n - 2 \left| c \right|.
\end{eqnarray}

\noindent Such different forms of compressibility reveal an important insight
into the root patterns of the quasi-momenta. We now show the
subtlety of the Bethe ansatz roots for the super Tonks--Girardeau
state\cite{31} in Fig.~\fref{f5}.

\vspace*{1mm}

\begin{center}
\fl{f5}\includegraphics[width=\linewidth]{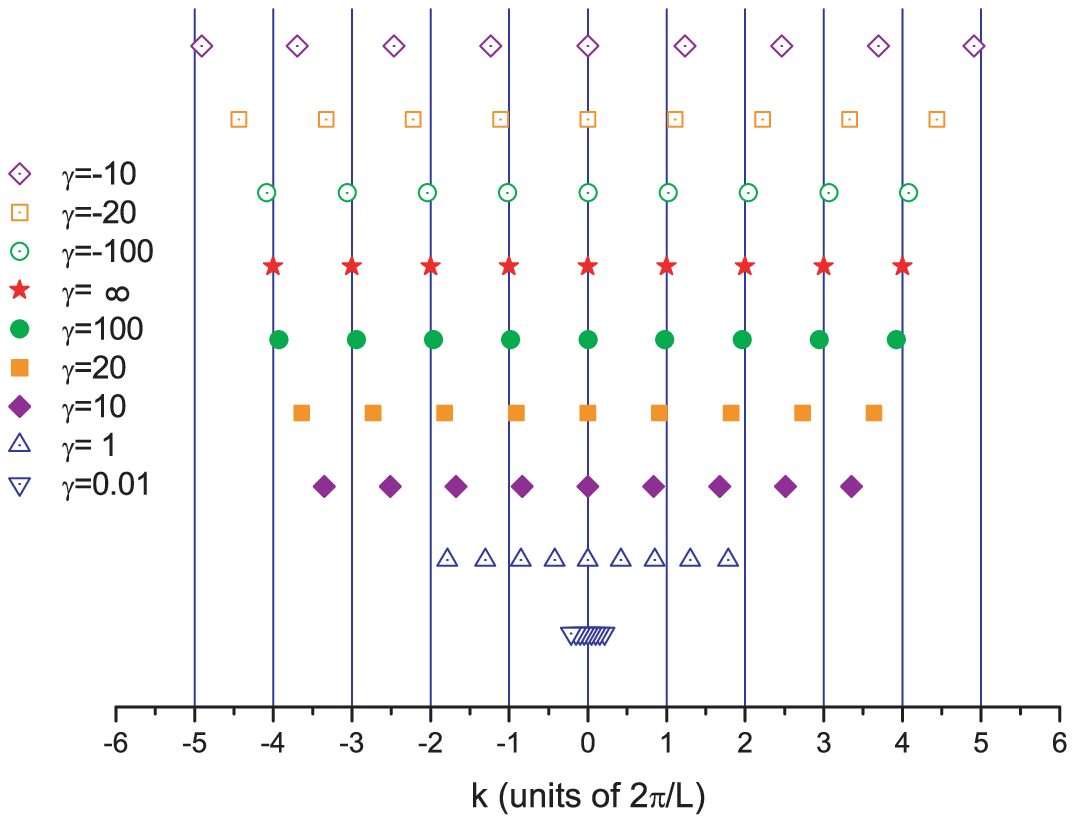}\\[5pt]
\parbox[c]{8.0cm}{\footnotesize{\bf Fig.~5.}
Quasi-momenta for the ground state of the Lieb--Liniger gas with the
strong repulsion and the supper Tonks--Girardeau gas  of the
attractive Bose gas for a large  value of $\gamma$. It is obvious
that the super Tons--Girardeau gas  has a larger kinetic energy than
the free Fermion momentum. Thus the quantum statistics of the super
Tonks--Girardeau gas is more exclusive than the free fermion
statistics.}
\end{center}

\subsection{Luttinger parameter and correlation functions}

The 1D integrable system gives rise to the power law behavior of
long distance or long time asymptotics of correlation functions for the ground state.
The effective Hamiltonian can be approximately described by the conformal
Hamiltonian which can be written in terms of the generators of the underlying Virasoro
algebra with the central charge $C=1$.
The low-lying excitations  present  the  phonon dispersion $\Delta E(p) =v_s p$ in the long-wavelength limit.
In this limit, all particles participate in the excitations and form
a collective motion of bosons which is called the Luttinger
liquid.\cite{35}
The Lieb--Liniger field theory Hamiltonian can be rewritten as an  effective Hamiltonian in long-wavelength limit,  which essentially describes the low-energy physics of the Lieb--Linger Bose gas
\begin{eqnarray}\el{50}
H=\int \zd x \left(\frac{\pi v_{\rm s} K}{2} \Pi^2+\frac{v_{\rm
s}}{2\pi K }\left(\partial _x \phi \right)^2\right),
\end{eqnarray}
where the canonical momenta $\Pi$ conjugate to the phase  $\phi$  obeying  the standard Bose commutation relations $\left[ \phi(x), \Pi(y) \right]=\mathrm{i} \delta(x-y)$.
$\partial_x \phi$ is proportional to the density fluctuations.
In this effective Hamiltonian, $v_s/K$ fixes the energy for the change of density.
In this approach, the density variation in space is viewed as a superposition of harmonic waves.

For example, the leading order of one-particle correlation $\langle{
\psi^{\dag} (x) \psi(0) } \rangle\sim 1/x^{1 / 2 K}$ is uniquely
determined by the Luttinger parameter $K$.
The Luttinger parameter is defined by the ratio of sound velocity to stiffness, namely,
\begin{eqnarray}\el{51}
 K=\frac{v_{\rm s}}{v_N}
 = \frac{\pi}{\sqrt{3 e_0(\gamma) -2\gamma\frac{{\rm d}e_0(\gamma)}{{\rm d} \gamma} + \frac{1}{2} \gamma^2 \frac{{\rm d}^2  e_0(\gamma)}{{\rm d}
  \gamma^2}}},
\end{eqnarray}
where $v_{\rm s}$ is the sound velocity and $v_N$ is the stiffness and defined as
\begin{eqnarray}\el{52}
  v_{N} = \frac{L}{\pi \hbar} \frac{\partial^2 E}{\partial N^2},~~
  v_{\rm s} = \sqrt{\frac{L^2}{mN} \frac{\partial^2 E}{\partial L^2}}.
\end{eqnarray}
In Eq.~(\eref{51}), the second expression of the Luttiger parameter
is used for numerical calculation with the help of Eq.~(\eref{22}).

Using the asymptotic expansion result of the ground-state energy
(\eref{24}) for weak counting and (\eref{29}) for the strong
counting regimes, we find the asymptotic forms of the Luttinger
parameter $K$ for the two limits
\begin{eqnarray}\el{53}
&  &
  K|_{\gamma\ll1} = \pi \bigg( \gamma - \frac{1}{2 \pi} \gamma^{3 / 2} \bigg)^{-1 /2}, \\
&&\el{54}
  K|_{\gamma\gg1} = 1 + \frac{4}{\gamma} + \frac{4}{\gamma^2} - \frac{16 \pi^2}{3 \gamma^3}.
\end{eqnarray}
In Fig.~\fref{f6}, we show that these asymptotic forms of the
Luttinger parameters provide a very accurate expression throughout
the whole parameter space.
The correlation functions can be calculated by the conformal field
theory.\cite{11,35,36,37,38}

\vspace*{2mm}

\begin{center}
\fl{f6}\includegraphics[width=\linewidth]{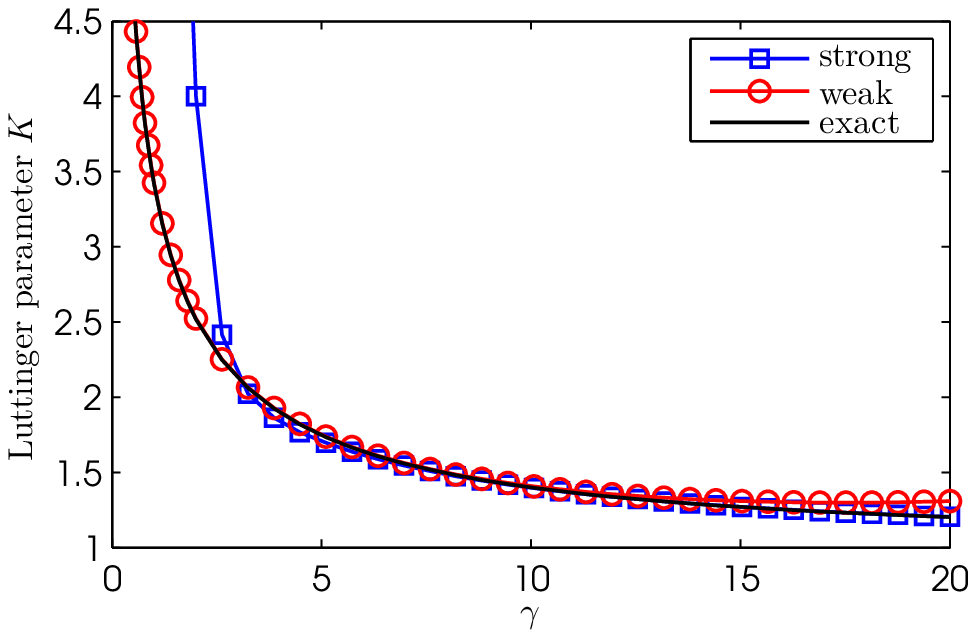}\\[5pt]
\parbox[c]{8.0cm}{\footnotesize{\bf Fig.~6.}
Luttinger parameter $K$ versus interaction strength  $\gamma$. Blue
dashed  line: the result obtained from
   Eq.~(\eref{54}) for the strong coupling limit; red dashed  line: result obtained from Eq.~(\eref{53}) for weak coupling regime;
   black solid line: the numerical result obtained from Eq.~(\eref{51}). The analytical
  result of the Luttinger parameter is in a good agreement with  the numerical result.}
\end{center}
\vspace*{2mm}

Usually, the long-distance or long-time asymptotics of correlation
functions of the 1D critical systems can be calculated by using  the
canonical field theory (CFT).\cite{11}
>From the CFT, the two-point correlation function for primary fields
with the conformal dimensions $\varDelta^{\pm}$ is given
by\cite{39}
\begin{eqnarray}\el{55}
G_{\rm O}(y,\tau) =\sum
 \frac{A {\rm e}^{-2\pi{\rm i}(N\Delta D)y/L}}{
 (v \tau+{\rm i}y)^{2\varDelta^+}
 (v \tau-{\rm i}y)^{2\varDelta^-}},
\end{eqnarray}
where $\tau$ is the Euclidean time, $v$ is the velocity of light,
$G_O(x,t)=\langle G|\hat O^{\dagger}(x,t)\hat O(0,0)|G\rangle $ is
the correlator for the field operators $\hat O(0,0)$ and  $\hat
O^{\dagger}(x,t)$.
Equation (\eref{55}) involves the contributions of the excited
states which are characterized by numbers $\Delta D$, $N^\pm$, and
$\Delta N$.
Here $N^-$ (or $N^+$) describes the elementary particle--hole
excitations by moving atoms close to the left (or right) pseudo
Fermi point  outside  the Fermi sea with adding $N^-$ (or $N^+$)
holes below the left (or right) Fermi point.
$\Delta N$ is  the change of particle number and it characterizes the elementary excitations by adding (or removing) particles over  the ground state.
$N^\pm$ and $\Delta N$ are not enough to describe all elementary excitations.
It is necessary to introduce the quantum number $2\Delta D$ to denote the particle number difference
between the right- and left-going particles.
>From the analysis of finite-size corrections of the BAE, the total momentum and excited energy of the low-lying  excitations in terms of these quantum numbers $N^\pm$, $\Delta N$, and $D$ are given by
\begin{eqnarray}\el{56}
 \Delta P&=&\frac{2\pi}{L}
 \big[\Delta N \Delta D +N^+-N^-\big]+2\Delta D k_{{\rm F}},\\
\el{57} \Delta E&=&\frac{2\pi v}{L}\bigg[
 \frac14(\Delta N/Z)^{2}
 +(\Delta D Z)^2+N^++N^-\bigg],
\end{eqnarray}
where $Z=2\pi\rho(Q)$ is the dressed charge at the pseudo Fermi point for the ground state.
>From the conformal field theory, the excited energy and momentum are given by
\begin{eqnarray}\el{58}
& &
 \Delta E=\frac{2\pi v}{L} v(\varDelta^++\varDelta^-),\\
 &&\el{59}
 \Delta P=\frac{2\pi}{L} \sum_\alpha (\varDelta^+-\varDelta^-)
 +2\Delta D k_{{\rm F}},
\end{eqnarray}
By comparison between the two results obtained from finite-size
corrections (\eref{56}), (\eref{57}), and CFT (\eref{58}),
(\eref{59}),  the conformal dimensions are analytically obtained as
a function of $N^{\pm}$, $\Delta N$, $\Delta D$, and the dressed
charge $Z$ as well, namely,
\begin{eqnarray}\el{60}
2\varDelta^\pm = 2N^\pm
 \pm\Delta N \Delta D
 +(\Delta D Z)^2
 +\frac{1}{4}(Z^{-1} \Delta N)^2.
\end{eqnarray}

It turns out that only the low-energy excitations determine the long distance or time asymptotic behavior of the correlation functions.
Given that the operator $\hat \psi(x)$ is a prime operator of this $U(1)$ symmetric system, the correlation functions of the prime field have a universal power law decay in distance
\begin{eqnarray}\el{61}
 \langle {\hat \psi^\dag(\tau,y)\hat \psi(0,0)}\rangle
 =\frac{\exp(2\mathrm{i}\Delta Dk_{\rm F}y)}{(v\tau +\mathrm{i} y)^{2\varDelta^+}(v\tau -\mathrm{i} y)^{2\varDelta^-}}.
\end{eqnarray}
We find that the Luttinger parameter $K=Z^2$ by comparing with the
result of the Luttinger theory, $\langle{\hat \psi^\dag(y)\hat
\psi(0)}\rangle \sim 1/y^{1/2K} $.
For example, in the strong coupling regime, the dressed charge
$Z=1+2/\gamma-8\pi^2/3\gamma^3$ from the dressed charge
equation.\cite{11}
Submitting it into $K=Z^2$, we obtain the same result as
Eq.~(\eref{54}).

\section{Yang--Yang thermodynamics and quantum criticality}

In 1969, Yang C N and Yang C P presented a grand canonical ensemble
to describe finite-temperature thermodynamics for
 the Lieb--Linger model.\cite{3}
The Yang--Yang method has led to significant developments in quantum
integrable systems.\cite{12,13,40}
This approach allows one to access full finite temperature physics
of the models in terms of the thermodynamic Bethe ansatz  (TBA)
equations.\cite{12}
In the grand canonical ensemble, we usually convert the TBA equations in terms of the dimensionless chemical potential $\tilde\mu=\mu/c^2$ and the dimensionless temperature $\tilde T=T/c^2$ with the interaction strength $c$.
It is also convenient to use the degenerate temperature as an energy
unit, i.e.,  $T_{\rm d}=\hbar^2 n^{2}/2m$.\cite{41,42} It is very
insightful to discuss the critical phenomena of the models in terms
of the dimensionless units.
 We first discuss the Yang--Yang grand canonical ensemble below.

\subsection{The Yang--Yang grand canonical ensemble}

For the ground state, the set of quantum numbers (\eref{18})
provides the lowest energy.
However, at finite temperatures, any thermal equilibrium state involves many microscopic eigenstates.
As discussed in previous section, these eigenstates are
characterized by different quantum numbers $\{I_j\}$, see
Eq.~(\eref{16}).
In the thermodynamic limit, $I(k)$ is a monotonic function of pseudo
momenta $k$.\cite{11}
We define ${\rm d}I(k)/(L{\rm d}k)=\rho(k) +\rho_{\rm h}(k)$, where $\rho_{\rm h}$ is the density of the holes.
>From Eq.~(\eref{16}), we can obtain the integral BAE for arbitrary
eigenstate as
\begin{eqnarray}\el{62}
 \label{ibae}
  \rho (k) + \rho_{\rm h} ( k) = \frac{1}{2 \pi}+\int_{- \infty}^{\infty} a(k-k') \rho (k') {\rm d} k'.
\end{eqnarray}
Here we should notice that integral interval can extend to the whole real axis, i.e.,
particles can occupy any real quasimomentum.

In order to understand the equilibrium states of the model, it is
essential to introduce the entropy.
In a small interval ${\rm d} k$, the number of total vacancies is $L [ \rho ( k) + \rho_{\rm h} (k)] {\rm d} k$ with a number of  $L\rho (k){\rm d} k$ particles and a number of $L\rho_{\rm h} (k) {\rm d} k$ holes.
These particles and holes give rise to microscopic states
\begin{eqnarray}
  {\rm d} W = \frac{[L(\rho ( k) + \rho_{\rm h} ( k)) {\rm d} k] !}{[ L \rho ( k) {\rm d} k] ! [ L
  \rho_{\rm h} (k) {\rm d} k] !}.\nonumber
\end{eqnarray}
In the thermodynamic limit, $[L(\rho ( k) + \rho_{\rm h} ( k)) {\rm d} k]\gg 1$ and ${\rm d}k\to 0$, with the help of Stirling's formula, the entropy in this small interval is given by
\begin{eqnarray}
 {\rm d} S=\ln {\rm d} W\approx
  L\{\rho\ln[1+ \eta]
 +\rho_{\rm h}\ln[1+ \eta^{-1}]\} {\rm d} k,
 \nonumber
\end{eqnarray}
where $\eta(k)=\rho_{\rm h}(k)/\rho(k)$.
The total entropy is $S=\int {\rm d}S$.
It can be understood from this procedure that the entropy involves the disorder of mixing the particles and holes in the  pseudo momentum space.
For regions with zero $\rho(k)$ or zero $\rho_{\rm h}(k)$, no disorder occurs, i.e.,  ${\rm d}S=0$.
Therefore, the entropy for the ground state is zero.
It is worth noting that the above discussion is valid only in the equilibrium state.

The Gibbs free energy $\varOmega$ is the thermodynamic potential of
grand canonical ensemble for this model,
\begin{eqnarray}\el{63}
 \varOmega=E-TS-\mu N,
\end{eqnarray}
where $\mu$ is the chemical potential, and the particle number is
given by $N=L\int \rho(k) {\rm d}k$. In thermal equilibrium, the
true physical state is determined by the conditions of minimizing
the Gibbs free energy. Making a virtual change $\delta \rho, \delta
\rho_{\rm h}$ in the thermal equivalent state, we take the variation
of the free energy such that
\begin{eqnarray}
  \delta \varOmega = \delta E - T \delta S - \mu \delta N = 0.
  \nonumber
\end{eqnarray}
Here  we should notice that the variations $\delta \rho$ and $\delta
\rho_{\rm h}$ are not independent in view of the integral
BAE~(\eref{62}).
This minimization condition leads to the TBA equations in terms of
the dressed energy\cite{3}
\begin{eqnarray}\el{64}
  \varepsilon(k) = k^2 - \mu + \int_{- \infty}^{\infty}a(k-k')\varepsilon_-(k'){\rm d}k',
\end{eqnarray}
where the dressed energy is defined by $\varepsilon(k) = T\ln\eta(k)$.
In the above equation, we also denote $\varepsilon_-(k) =-T\ln[1+{\rm e}^{-\varepsilon(k)/T}]$.
Using the TBA equation~(\eref{64}), we further obtained   the grand
thermodynamic potential  $ \varOmega = \frac{L}{2 \pi} \int
\varepsilon_-(k){\rm d} k$.
It follows that the pressure is given by
\begin{eqnarray}\el{65}
  p=-\bigg(\frac{\partial \varOmega}{\partial L} \bigg)_{\mu,c,T}=-\frac{1}{2 \pi} \int \varepsilon_-(k){\rm d} k.
\end{eqnarray}
This serves as the equation of state (\eref{65}) from which we can
calculate the thermodynamics of this model at finite temperatures.
We can obtain the zero-temperature  and finite-temperature phase diagrams of the Lieb--Liniger model.
In  the zero-temperature limit, the TBA  equation~(\eref{64})
reduces to the dressed energy equation (\eref{38}) in the limit
$T\to 0$. From the standard thermodynamic relations, one can
calculate the particle density $n =\partial_\mu p|_{c, T}$, entropy
density $s=\partial_T p|_{\mu, c}$, compressibility
$\kappa^*=\partial^2_\mu p|_{c, T}$, specific heat $c_{\rm
v}=T\partial^2_T p|_{\mu,c}$ in a straightforward way.

\vspace*{2mm}

\begin{center}
\fl{f7}\includegraphics[width=\linewidth]{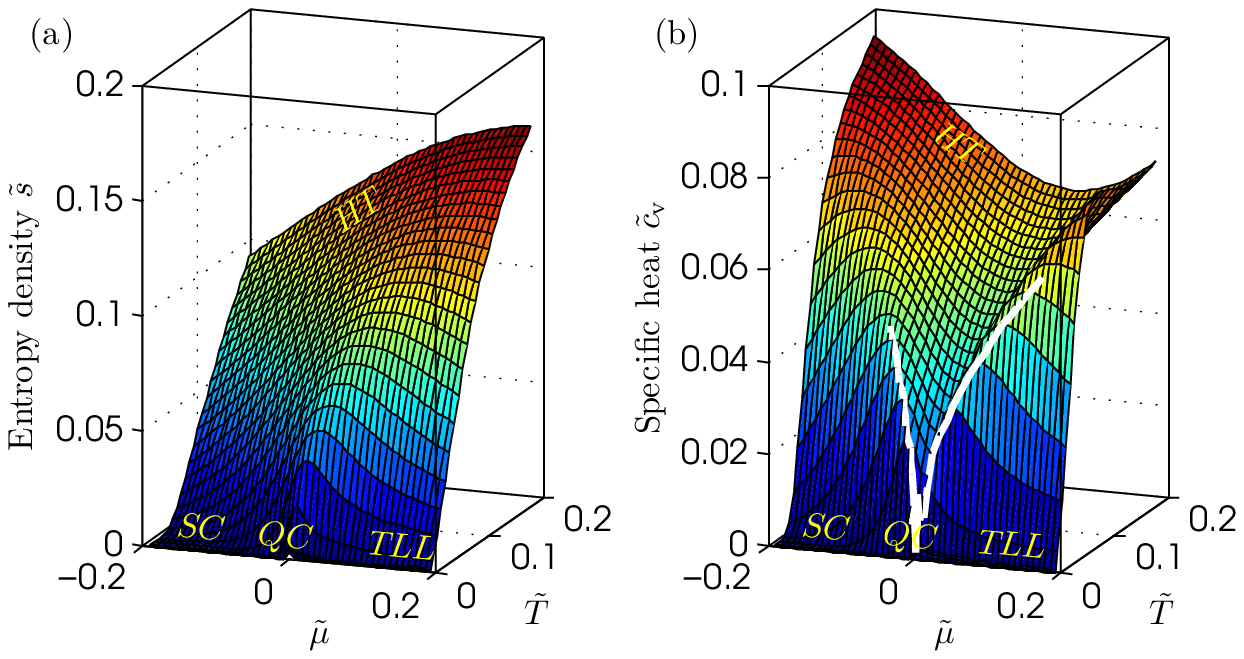}\\[5pt]
\parbox[c]{8.0cm}{\footnotesize{\bf Fig.~7.}
  Quantum critical regimes for the Lieb--Liniger model.
  (a), (b) Dimensionless entropy and specific heat in the $\tilde T$--$\tilde\mu$ plane, respectively.  For $\tilde T\gg1$, it is the HT regime.
  The TLL with the dynamical exponent $z=2$ and correlation length exponent $\nu =1$ lies in the region $\tilde T\ll1$ and $\mu> \mu_{\rm c}$.
For  $\tilde T\gg |\mu-\mu_{\rm c}|$, the QC regime with $z=2$, $d=1$, and $\nu=1/2$ fans out near the quantum phase transition point $\tilde \mu_{\rm c}=0$.
It is obvious that the entropy and the specific heat have
singularity properties near the critical point $\mu_{\rm c}=0$.}
\end{center}

In view of the grand canonical ensemble, there exists a quantum
phase transition at the chemical potential $\mu_{\rm c}=0$ at zero
temperature. Universal thermodynamics is expected for the
temperature under the quantum degenerate regime $T\lesssim T_{\rm
d}$, where $T_{\rm d}=\hbar^2 n^{2}/2m$.\cite{41,42}
At high temperatures (HT), i.e., $T\gg T_{\rm d}$, the system behaves like a classical Boltzmann gas.
For $\mu<\mu_{\rm c}=0$ and at low temperatures, the density is very low and the gas becomes de-coherent.
This phase is semiclassical (SC).
Whereas for $\mu>\mu_{\rm c}$ and the temperature $T<|\mu-\mu_{\rm
c}|$, it shows the Tomonaga--Luttinger liquid (TLL)  phase.
The quantum critical regime lies between SC and TLL for the temperature  $T\gg |\mu-\mu_{\rm c}|$.
We show such different regimes through the entropy and specific heat
in Fig.~\fref{f7}.

\subsection{The Yang--Yang equation and quantum statistics}

Dynamical interaction and thermal fluctuation drive Lieb--Liniger model from one phase into another.
In particular, under the degenerate temperature $T_{\rm d}$, the model has three distinct phases: semi-classical, quantum critical, and the TTL critical phases.
At high temperatures, there does not exhibit universal behaviour.
When the temperature tends to be infinity, the system reaches the Boltzmann gas.
Therefore, the Yang--Yang equation~(\eref{64}) encodes different
quantum statistics.
For example, when the coupling strength $\gamma\to 0$, the system behaves as the free bosons;
for the strong coupling limit, it behaves like free fermions;
at high temperatures, the system becomes the Boltzmann gas.
In the following, we rigorously derive such quantum statistics in an analytical way.
When the coupling strength $\gamma$ turns to be zero, the integral kernel $a(k)\to \delta(x)$.
The dressed energy in this limit can be expressed as
\begin{eqnarray}\el{66}
 \lim_{c\to0}\varepsilon(k)=T\ln\big[{\rm e}^{{(k^2-\mu)/T}}-1\big],
\end{eqnarray}
from which we obtain the thermal potential per length
$\lim_{c\to 0} \varOmega/L=\int_0^\infty 2\sqrt{\epsilon}/[{\rm e}^{(\epsilon-\mu)/T}-1] ~{\rm d}\epsilon$.
Thus, the distribution function satisfies the Bose--Einstein statistics
\begin{eqnarray}\el{67}
 \lim _{c\to 0}g(\epsilon)=\frac1{{\rm e}^{(\epsilon-\mu)/T}-1}.
\end{eqnarray}
If $\gamma\to\infty$, the dressed energy reads
\begin{eqnarray}\el{68}
 \lim_{c\to\infty}\varepsilon(k)=k^2-\mu.
\end{eqnarray}
In this limit, the Bethe ansatz equation (\eref{62}) naturally
reduces to the form
\begin{eqnarray}\el{69}
\rho(k) =\frac{1}{2\pi(1+\rho_{\rm h}(k)/\rho(k)) },
\end{eqnarray}
which indicates the Fermi--Dirac statistics.
Consequently, the thermal potential per unit length is given by
$\lim_{c\to \infty} \varOmega/L=\int_0^\infty
2\sqrt{\epsilon}/[{\rm e}^{(\epsilon-\mu)/T}+1]~{\rm d}\epsilon$.
This result gives rise to the Fermi--Dirac statistics
\begin{eqnarray}\el{70}
 \lim _{c\to\infty}g(\epsilon)=\frac1{{\rm e}^{(\epsilon-\mu)/T}+1}.
\end{eqnarray}

We further remark that equations (\eref{66})--(\eref{70}) are valid
for arbitrary temperature.
If the system is under the quantum degeneracy, the quantum
statistical interaction is important. Thus, the particles are
indistinguishable.
At high temperatures, the Yang--Yang equation (\eref{64}) gives rise
to the Maxwell--Boltzmann statistic such that the particles are
distinguishable.
In the weak coupling limit and high-temperature limits, it is very
convenient to consider Viral expansions with the Yang--Yang equation
(\eref{64}), namely,
\begin{eqnarray}\el{71}
\e^{-\epsilon(k)/T}={\cal
Z}\e^{-{k^2}/{T}}\e^{\int_{-\infty}^{\infty} \zd q
a(k-q)\ln\left(1+{\cal Z} \e^{-{q^2}/{T}} \right)}.
\end{eqnarray}
Here ${\cal Z}=\e^{\mu/T}$ is fugacity. After some algebra, we find
that the pressure up to the second Viral coefficient is given by
\begin{eqnarray}\el{72}
p=p_0+\frac{T^{{3}/{2}}}{\sqrt{2\pi} }{\cal Z}^2p_2,
\end{eqnarray}
where $p_2=-\frac{1}{2}+\int_{-\infty}^{\infty}\zd q' a(2q')
\e^{-{2q'^2}/{T}}$ reveals the two-body interaction effect.
In the above equation, $p_0=-\frac{T}{2\pi}\int_{-\infty}^{\infty}
\zd k \ln \left(1-{\cal Z}\e^{-{k^2}/{T}} \right)$ is the pressure
of the free bosons.
The result (\eref{72}) gives the Maxwell--Boltzmann statistics in
the limit of $T\to \infty$.

It is remarkable to discover the universal low temperature behavior of the Lieb--Liniger model with the Bose--Einstein statistic and Fermi--Dirac statistic.
The Yang--Yang equation (\eref{64}) provides full physics of the
Lieb--Liniger model which goes beyond that can be found by
Bose--Fermi mapping.\cite{18}
In fact, for strong coupling limit, i.e.,  $\gamma\gg1$,  the system
can be viewed as an ideal gas with the fractional
statistics.\cite{43}
When  the coupling strength is very weak, the ground state behaves like a quasi BEC.
The Bogoliubov approach is valid in the weak coupling limit $\tilde
T \ll \sqrt{\gamma}\ll1$.\cite{27,41,42}
The result~(\eref{72}) is also a good approximation for the weak
coupling Lieb--Liniger gas.

\subsection{Luttinger liquid and quantum criticality}

\subsubsection{Equation of state}
In low-energy physics, $T\ll T_{\rm d}$, low-lying excitations form a collective motion of bosons.
The linear relativistic dispersion near the Fermi points results in the TLL behavior.
At finite temperatures, the TLL can be sustained in a region of
$T<|\mu-\mu_{\rm c}|$ in the $T$--$\mu$ plane (see Fig.~\fref{f7}).
In the TLL phase, we can take the Sommerfeld expansion with the TBA
equation (\eref{64}).
By iterations, the pressure with the leading order temperature correction is given by
\begin{eqnarray}\el{73}
 p=p_0+\frac{\pi^2T^2}{3}\frac{\rho_0(Q)}{\varepsilon'(Q)}
 =p_0+\frac{\pi T^2}{6v_{\rm s}},
\end{eqnarray}
which  gives the free energy per unit length as the field theory prediction
\begin{eqnarray}\el{74}
  F(T)/L = \mu n - p \approx E_0 - \frac{\pi C ( k_{\rm B} T)^2}{6 \hbar v_{\rm s}},
\end{eqnarray}
with the central charge $C=1$. Here we took  $k_{\rm B}=1$,
so that in TLL phase the specific heat is linear temperature-dependent
\begin{eqnarray}\el{75}
  c_{\rm v} = \frac{\pi T}{3 v_s}.
\end{eqnarray}
This is a universal signature of the TLL.

However, for the temperature beyond the crossover temperature, i.e.,
$T>T^{*} \sim |\mu-\mu_{\rm c}|$, the excitations give a
non-relativistic dispersion, i.e., $\Delta E \sim p^2$.
The crossover temperature $T^*$ can be also determined by the
breakdown of linear temperature-dependent relation given by
Eq.~(\eref{75}).\cite{25}
The crossover is also evidenced by the correlation length.\cite{44}

For  strong coupling and low temperatures, i.e., $\gamma\gg1$ and
$\tilde T\ll1 $, the pressure is given by\cite{25}
\begin{eqnarray}\el{76}
  p = - \frac{T^{3 /2}}{2 \sqrt{\pi}}{\rm Li}_{3/2}
  \big(-{\rm e}^{A/T}\big) \bigg[1+\frac{T^{3/2}}{2 \sqrt{\pi} c^3}
  {\rm Li}_{3/2}
  \big(-{\rm e}^{A /T}\big) \bigg],~~~~
\end{eqnarray}
where $A = \mu + {2 p}/{c} + ({T^{5 / 2}}/{2 \sqrt{\pi} c^3}) {\rm
Li}_{3/2} ( -{\rm e}^{A / T})$ and the polylogarithm function is
given by ${\rm Li}_n ( x) = \sum_{k = 1}^{\infty} \frac{x^k}{k^n}$.
In the above equation, the pressure $p$ gives a close form of the
equation of state.
Using the standard thermodynamical relations, we can analytically calculate the  particle density, compressibility, and the specific heat
\begin{eqnarray}\el{77}
  n &=& -\frac{1}{2 \sqrt{\pi}} T^{1/2} f_{1/2} \bigg\{ 1 -
  \frac{1}{\sqrt{\pi} c} T^{1/2} f_{1/2} + \frac{T}{\pi c^2}
  f_{1/2}^2 \notag\\
  &&+\, \frac{1}{\sqrt{\pi} c^3}
  T^{3/2} \bigg[ -
  \frac{1}{\pi} f_{1/2}^3 + \frac{3}{2} f_{3/2} \bigg]
  \bigg\},\\
 \el{78} \kappa^* &\approx& - \frac{1}{2 \sqrt{\pi}} T^{-1/2}
  f_{-1/2}
  + \frac{3}{2 \pi c} f_{-1/2} f_{1/2} \notag\\
  &&-\, \frac{2}{\pi^{3 / 2}
  c^2} T^{1/2} f_{-1/2} f_{1/2}^2 -
  \frac{1}{\pi c^3}
  Tf_{-1/2} f_{3/2} \notag\\
  &&+\, \frac{5}{\pi^2 c^3} Tf_{-1/2}
  f_{1/2}^3 - \frac{3}{4 \pi c^3} Tf_{1/2}^2,\\
\el{79}\frac{c_{\rm v}}{T} &=&  \bigg( \frac{\partial s}{\partial T}
\bigg)_{\mu, c} = - \frac{3}{8
  \sqrt{\pi}} T^{-1/2} f_{3/2} + \frac{1}{2 \sqrt{\pi}} T^{-1/2
}  \frac{A}{T} f_{1/2} \notag\\
  &&-\, \frac{1}{2 \sqrt{\pi}} T^{-
  \frac{1}{2}} \bigg( \frac{A}{T} \bigg)^2 f_{-1/2} + O \bigg(
  \frac{1}{c} \bigg),
\end{eqnarray}
respectively. Here $f_n = {\rm Li}_n ( - {\rm e}^{A / T})$.

\subsubsection{Quantum criticality}

At zero temperature, the quantum phase transition from the vacuum
phase into the TLL at the critical point $\mu_{\rm c}=0$ occurs in
the Lieb--Liniger Bose gas.
According to the renormalized group theory, universal scaling
properties are expected in the critical regime at low temperatures
(see Fig.~\fref{f7}).
In 2011, Guan and Batchelor investigated the quantum criticality of
the Bose gas and  found that the equation of state (\eref{76})
reveals the universal scaling behavior of quantum criticality in
terms of the polylogarithm functions.\cite{25}
It is straightforward from the equation of the state to derive the universal scaling form of the density as
\begin{eqnarray}\el{80}
  n(T, \mu) \approx n_0 + T^{d/z + 1 - 1/\nu z} \mathcal{F}
  \bigg(\frac{\mu - \mu_{\rm c}}{T^{1/\nu z}} \bigg),
\end{eqnarray}
where the background density $n_0 = 0$.
The scaling function $\mathcal{F}(x) = - \frac{1}{2 \sqrt{\pi}} {\rm
Li}_{1/2} (-{\rm e}^x)$ read off the dynamic critical exponent $z =
2$, and the correlation length exponent $\nu = 1 / 2$.
It is particularly interesting that the finite temperature density profiles of the 1D trapped gas can map out the quantum criticality with these universal exponents for the Lieb--Liniger gas.
The density curves at different temperatures intersect at the
critical point, see Fig.~\fref{f8}{(a)}.
The universal scaling behavior of compressibility $\kappa^*$ is given by
\begin{eqnarray}\el{81}
  \kappa^{*} = \kappa_0 + T^{d/z + 1 - 2/{\nu z}} \mathcal{K} \bigg(
  \frac{\mu - \mu_{\rm c}}{T^{1/\nu z}} \bigg),
\end{eqnarray}
where $\kappa_0 = 0$ and ${\cal Q}( x) = - \frac{1}{2 \sqrt{\pi}} {\rm Li}_{-
{1}/{2}} (x)$.
This scaling function again reads out the dynamic critical exponent $z = 2$, and the
correlation length exponent $\nu = 1 / 2$.
The intersection at the critical point for different temperatures
attributes to the universal scaling form (\eref{81}) (see
Fig.~\fref{f8}{(b)}).


\begin{center}
\fl{f8}\includegraphics[width=\linewidth]{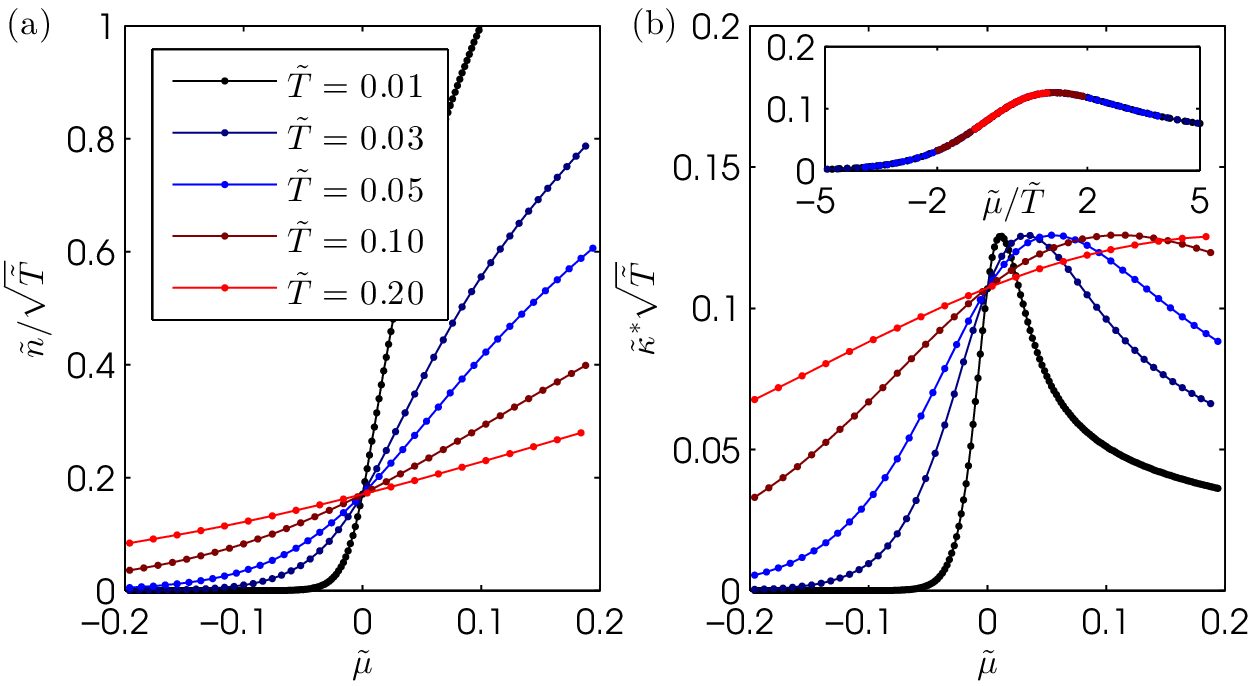}\\[5pt]
\parbox[c]{8.0cm}{\footnotesize{\bf Fig.~8.}  Universal scaling behaviors  of the density and compressibility at quantum criticality.  (a) Density shows the universal scaling behavior given by (\eref{80}).  (b) Compressibility presents the universal scaling
behavior~(\eref{81}).  The inset in (b) shows the collapse of
temperature-rescaled compressibility  $\tilde\kappa^*\sqrt{\tilde
T}$ with respect to the argument $(\tilde{\mu}-\tilde{\mu_{\rm
c}})/\tilde{T}$.}
\end{center}


\begin{center}
\fl{f9}\includegraphics[width=\linewidth]{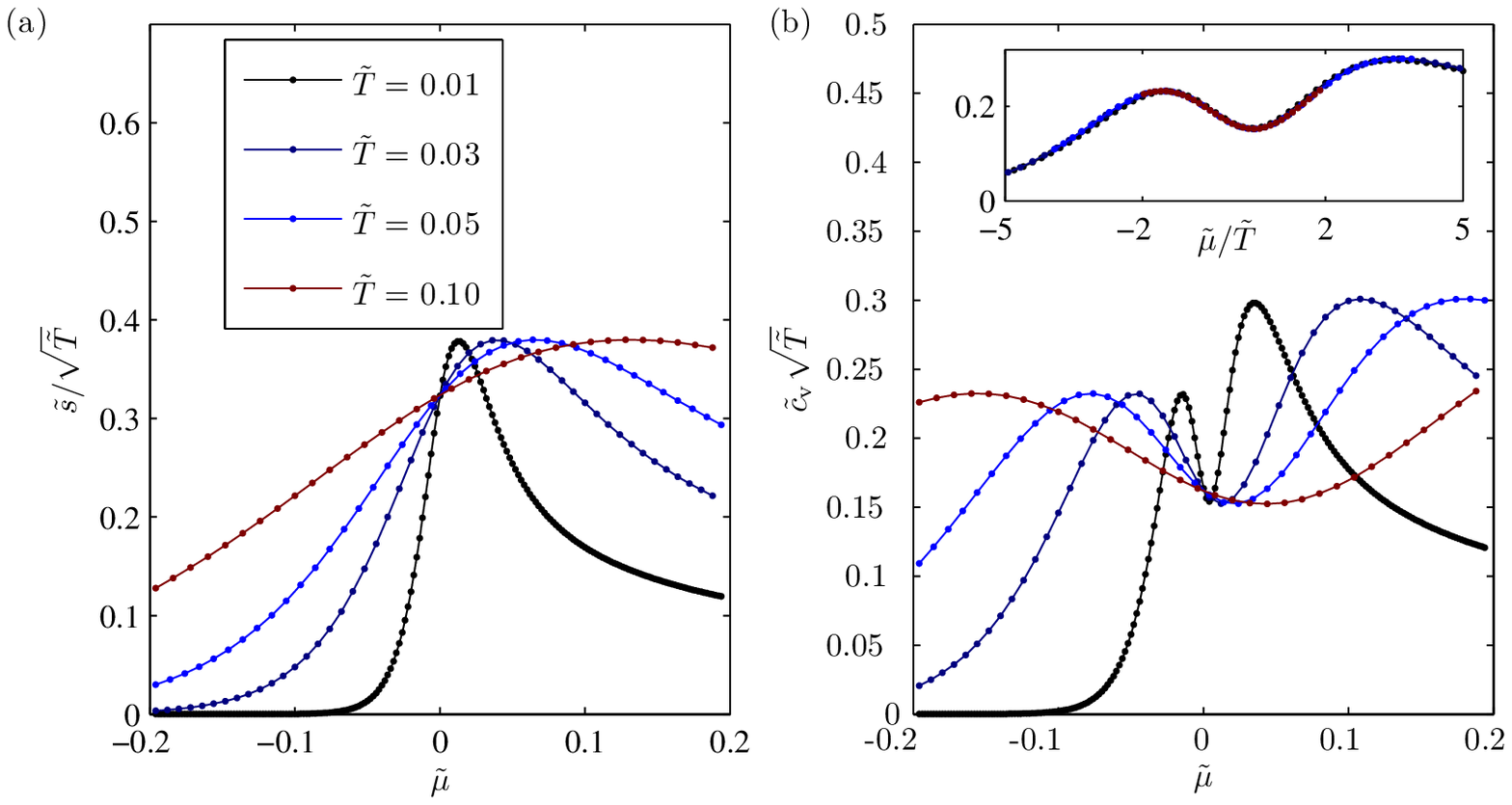}\\[5pt]
\parbox[c]{8.0cm}{\footnotesize{\bf Fig.~9.}
Quantum critical behaviour of the  entropy and specific heat.
 (a) The entropy divided by temperature $\tilde{s}=s/c$  has a universal  scaling bahavior at the quantum criticality.  (b) Specific heat divided by temperature $\tilde{c}_{\rm v} =c_{\rm v}/T$  has a universal  scaling bahavior at the quantum criticality.
 The inset in (b) shows the collapse of $\tilde c_{\rm v}\sqrt{\tilde T}$ with respect to the argument $(\tilde{\mu}-\tilde{\mu_{\rm c}})/\tilde{T}$.}
\end{center}
\vspace*{2mm}

Near the critical point, the specific heat divided by the
temperature $\tilde{c}_v\equiv c_{\rm v}/T$ obeys the following
scaling form:
\begin{eqnarray}\el{82}
  \tilde{c}_{\rm v}& =& -T^{-1/2}
  \bigg[\frac{3}{8\sqrt{\pi}}{\rm Li}_{3/2}\big(-{\rm e}^{\mu/T} \big) - \frac{1}{2\sqrt{\pi}} \frac{\mu}{T}{\rm Li}_{1/2}
  \big(-{\rm e}^{\mu/T} \big) \nonumber\\
&  & + \,\frac{1}{2 \sqrt{\pi}}  \bigg( \frac{\mu}{T}
  \bigg)^2 {\rm Li}_{-1/2}\big(-{\rm e}^{\mu/T}\big)
  \bigg].
\end{eqnarray}
The specific heat at different temperatures has two round peaks near
the critical point $\mu_{\rm c} = 0$.
These peaks mark the crossover temperatures that
distinguish the TLL and semi-classical gas phases from the quantum critical regime.
This is a very robust signature for the existence of the crossover temperatures in the 1D Bose gas.
This scaling law of the entropy and specific heat is shown in
Fig.~\fref{f9}.

\subsection{The local pair correlations and Tan's contact}

In the study of the interacting Bose gas in one dimension, an important property is the local two-body correlation function $g_2$.
Physically speaking, this function describes the rates of inelastic
collision between pairs of particles.\cite{45,46,47}
This quantity reflects the probability that the two particles site at the same size.
It is also known as contact that strikingly captures the universality of ultracold atoms.
This has been described by Tan's
relations.\cite{48,49,50}
Tan's contact, which measures the two-body correlations at short
distances in dilute systems, is a central quantity to ultra-cold
atoms.
It builds up universal relations among thermodynamic quantities such
as the large momentum tail, energy, and dynamic structure factor,
through the renowned Tan's relations, see recent developments in
this research.\cite{51}

Knowing that the two-body correlation can lead to the classification of physically distinct regimes, for example, the Tonks--Girardeau
regime with $g_2\to 0$, the Gross--Pitaevskii regime with $g_2=1$ and the very weak coupling or fully decoherent regime with $g_2=2$.
The two-body correlation function can be calculated for these
different regimes by expression $g_2=\langle \hat{\varPsi}
^{\dagger}(x)\hat{ \varPsi}^{\dagger }(x) \hat{\varPsi}(x)
\hat{\varPsi}(x)\rangle $.
where $\hat{\varPsi}$ is the field operator in second quantization.
At $T=0$, we have ${\rm d} E_0/{\rm d} c=L g_2$, where $E_0$ is the ground-state energy.
For weak coupling limit, the local pair correlation
$g_2/n^2=1-{2\sqrt{\gamma} }/{\pi}$.
For strong coupling limit,
$g_2/n^2={4\pi^2}/{3\gamma^2}\left(1-{6}/{\gamma} \right)$.
In general, the local pair correlation can be used to study phase coherence behavior at finite temperatures.
Introducing the free energy per particle, i.e., $f(\gamma, T) =F/N$, the normalized two-particle local correlation is defined as
\begin{eqnarray}\el{83}
g_2 = \frac{\langle\hat{ \varPsi }^{\dagger}(x)
\hat{\varPsi}^{\dagger }(x) \hat{\varPsi}(x)
\hat{\varPsi}(x)\rangle}{n^2}=\frac{2m}{\hbar^2 n^2}\left(
\frac{\partial f(\gamma, T) }{\partial \gamma } \right)|_{n,T}.~~~~~
\end{eqnarray}
For strong coupling regime, the local pair correlation function is
obtained from the equation of state\cite{46,47,52,53}
\begin{eqnarray}\el{84}
g_2=\frac{4\pi^2}{3\gamma^2}\left(1-\frac{6}{\gamma}+ \frac{T^2}{4\pi^2 T_d^2} \right).
\end{eqnarray}

On the other hand, in one dimension the fundamental thermodynamic
relation in a harmonic trap is given by\cite{51}
\begin{eqnarray}\el{85}
  {\rm d} p = n{\rm d}\mu + s{\rm d}T-\frac{\rho_{\rm s}}{2} {\rm d} w^2 - \mathcal{C} {\rm d} a_{\rm
  1D},
\end{eqnarray}
where $\rho_{\rm s}$ and $\mathcal{C}$ are the densities of superfluid and contact, respectively.
In this relation, $w=v_{\rm s}-v_{\rm n}$ is the difference between the velocity of the superfluid and normal components.
Maxwell relations build the general connections between the contact and other physical quantities such as
\begin{eqnarray}
\bigg( \frac{\partial \mathcal{C}}{\partial \mu} \bigg)_{T, a_{\rm 1
D}}& =& -
  \bigg( \frac{\partial n}{\partial a_{\rm 1 D}} \bigg)_{\mu, T},\nonumber \\
 \bigg( \frac{\partial \mathcal{C}}{\partial T} \bigg)_{\mu, a_{\rm 1 D}} &=&-
  \bigg( \frac{\partial s}{\partial a_{\rm 1 D}} \bigg)_{\mu, T}.\nonumber
\end{eqnarray}
Furthermore, we can obtain the contact through the following relations:
\begin{eqnarray}\el{86}
  \mathcal{C} &=& - \frac{1}{c^2} \bigg( \frac{\partial p}{\partial c}
  \bigg)_{\mu, T} \approx \frac{1}{2 \pi} T^2 f_{1/2}
  f_{3/2}\notag\\
  &&\times\,
  \bigg( 1 - \frac{1}{\sqrt{\pi} c} T^{1/2} f_{1/2}
  \bigg).
\end{eqnarray}
The Tan's contact of the Lieb--Liniger gas does not have the usual
scaling behavior which was found for the interacting Fermi gas in
Ref.~\cite{51}.
This is mainly because the critical field $\mu_{\rm c} = 0$ which
does not depend on the scattering length $a_{\rm 1 D}$ (see
Fig.~\fref{f10}).

\vspace*{2mm}

\begin{center}
\fl{f10}\includegraphics[width=\linewidth]{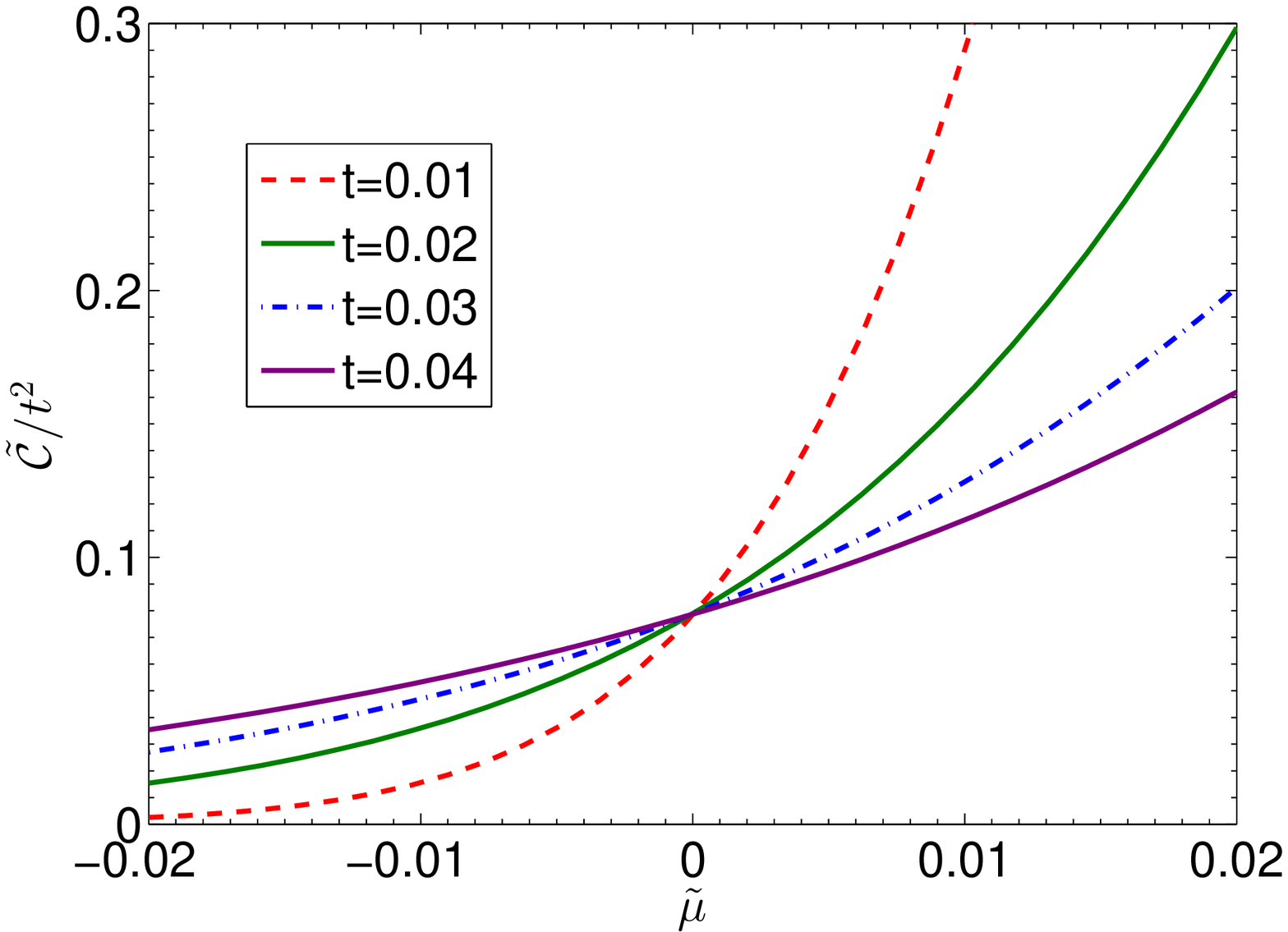}\\[5pt]
\parbox[c]{8.0cm}{\footnotesize{\bf Fig.~10.}
  Contact divided by $\tilde T^2$ versus chemical potential.
   The critical scaling behavior of the contact for the Lieb--Liniger
gas is different from that for the 1D interacting Fermi
gas.\cite{27}}
\end{center}

\section{Experimental development related to the Lieb--Liniger gas}

Over the past few decades, experimental achievements in trapping and cooling ultra-cold atomic gases have revealed beautiful physics of the cold quantum world.
In particular, recent breakthrough experiments on trapped ultracold bosonic and fermionic atoms confined to one dimension have provided a precise understanding of significant quantum statistical and strong correlation effects in quantum many-body systems.
The particles in the waveguides are tightly confined in two transverse directions and weakly confined in the axial direction.
The transverse excitations are fully suppressed by the tight confinements.
Thus, the atoms in these waveguides can be effectively characterised
by a quasi-1D system. Thus, 1D effective  interaction potentials can
be controlled in the whole interacting regime by the underlying 3D
scattering with tight confinements in the two transverse
directions.\cite{16,16b,17}
 In such a way, these 1D many-body systems ultimately relate to the integrable models of interacting bosons and  fermions.
It is now possible to realize effectively one-dimensional quantum
Bose gases, in which the interaction strength between ultracold
atoms is tunable, see recent reviews.\cite{36,40}
These experiments have successfully demonstrated the anisotropic
confinements of atoms to one dimension by optical waveguides, see a
feature review article.\cite{54}
Particularly striking examples involve the measurements of momentum
distribution profiles,\cite{23,51}  the ground state of the
Tonks--Girardeau gas,\cite{24} quantum
correlations,\cite{55,56,57,58,59,60}  Yang--Yang
thermodynamics,\cite{61,62}  the super Tonks--Girardeau
gas,\cite{63} quantum phonon fluctuations,\cite{64,65,66}
elementary excitations and dark solitons,\cite{67,68}
thermolization and quantum dynamics.\cite{69,70,71} More
experimental developments of  the Lieb--Liniger model are listed  in
Table~\ref{t1}.\cite{72}

\label{t1}
\begin{center}
{ \footnotesize{\bf Table 1.} Experiments of Lieb--Liniger gas.}
\vspace{1.5mm}

\tabcolsep 18pt \footnotesize{
\renewcommand\arraystretch{1.2}

\begin{tabular}{ll}
\hline\hline\hline
quantum dynamics & $^{87}$Rb\cite{60,70,73,74,75}\\
  thermalization& $^{87}$Rb\cite{60,69,73,74}\\
  solitons & $^{87}$Rb\cite{67,76}\\
  fermionization& $^{39}$K\cite{56,71}\\
  YY thermodynamics& $^{87}$Rb\cite{55,60,61,64,65,66,77}\\
  strong coupling& $^{87}$Rb\cite{23,24}\\
  phase diagram & Cs\cite{78}\\
  3-body correlations  & $^{87}$Rb,\cite{57,62} Cs\cite{59}\\
  excited state & Cs\cite{63}\\
\hline\hline\hline
\end{tabular}}
\end{center}
\vspace*{2mm}

The early experimental studies of the Lieb--Linger gas with cold
atoms were made in the laboratory by Bloch's group\cite{23} and
Weiss's group.\cite{24}
In particular, the observation of the ground state energy of the
Tonks--Girardeau gas provides deep insights into understanding
fermionization effect induced by a strong repulsive interaction, see
Fig.~\fref{f11}.
Loading the ${}^{87}$Rb ultracold atoms into a 2D array of 1D tubes, where the atoms were kept in the lowest energy state in the two transverse directions.
Thus, the systems were realized in quasi-1D systems within axial
harmonic traps.
The essential feature of the Tonks--Girardeau gas was observed through the ground-state energy $T_{\rm 1D}$ of such waveguided ensembles.

\vspace*{2mm}

\begin{center}
\fl{f11}\includegraphics[width=\linewidth]{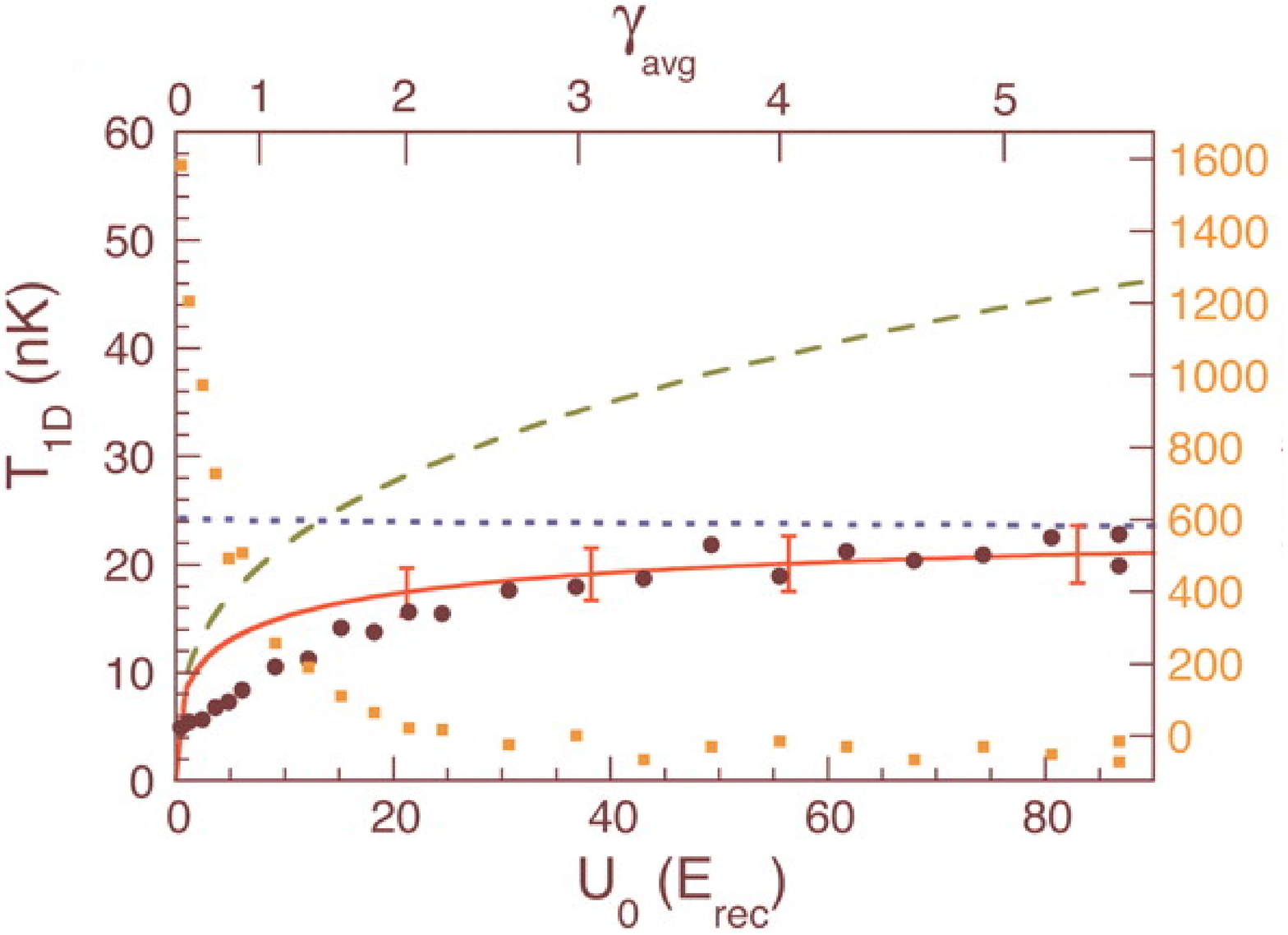}\\[5pt]
\parbox[c]{8.0cm}{\footnotesize{\bf Fig.~11.}
The 1D ground-state energy $T_{\rm 1D}$ versus transverse
confinement depth of the lattice. For the confinement potential
$U_0>0$ (or the effective interaction $\gamma\gg 1$), the  energy
$T_{\rm 1D}$ well presents the ground-state energy of the
Lieb--Liniger gas in strong coupling regime within the local density
approximation.\cite{16b}}
\end{center}
\vspace*{2mm}

The experimental measurement of the metastable highly excited state
--- the super Tonks--Girardeau gas was achieved by Haller\cite{63}
in 2009.
They  made a new experimental breakthrough with a stable highly
excited gas-like phase in the strongly attractive regime of bosonic
Cesium atoms across a confinement-induced resonance, see
Fig.~\fref{f12}.
This particular state was first predicted theoretically by
Astrakharchik {\em et al.}\cite{29} Using the Monte Carlo method
and by ANU group from the integrable interacting Bose gas with
attractive interactions.\cite{30}
This model has improved our understanding of quantum statistics and dynamical interaction effect in many-body physics.
It turns out that a highly excited state of gas-like gas could be
stable as the interaction is switched from strongly repulsive into
strongly attractive interactions due to the the existence of
Fermi-like pressure.\cite{31,32,33}
This phenomenon has triggered much attention in theory.\cite{79}

\vspace*{2mm}

\begin{center}
\fl{f12}\includegraphics[width=\linewidth]{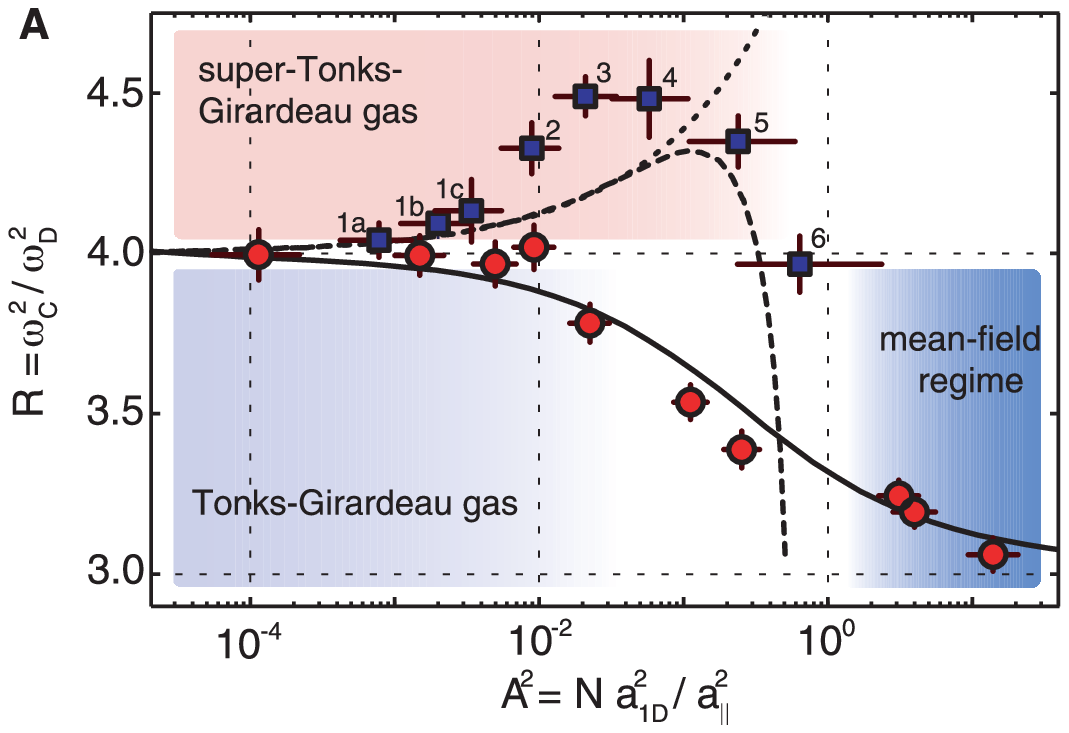}\\[5pt]
\parbox[c]{8.0cm}{\footnotesize{\bf Fig.~12.}
The ratio of compress mode over the trapping frequency
$R=\omega^2_{\rm c}/\omega^2 _{\rm D}$ versus the interaction
parameter $A^2$. The squares show the experimental  measurements in
strong attractive regime. The circles show the experimental data
ranging from weak coupling to strong TG regimes. The black solid
line stands for the exact result from the Lieb--Linger gas with a
repulsion. The dashed lines present the theoretical data from the
result.\cite{29} Further study of the Tonks--Girardeau gas can be
found in Ref.~\cite{31}.}
\end{center}
\vspace*{1mm}


\begin{center}
\fl{f13}\includegraphics[width=\linewidth]{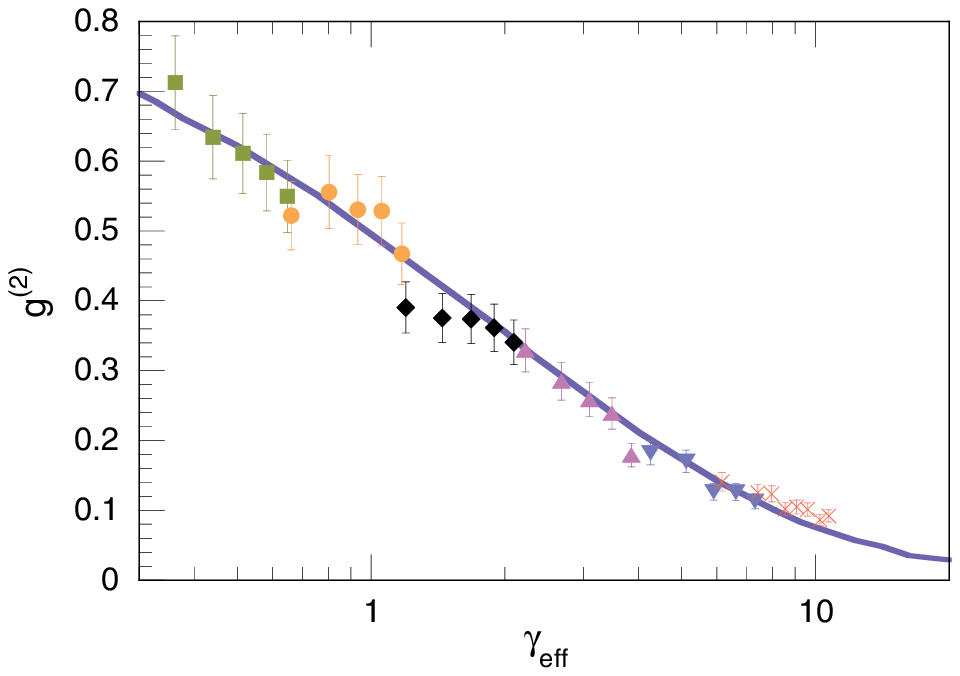}\\[5pt]
\parbox[c]{8.0cm}{\footnotesize{\bf Fig.~13.}
The local pair correlation function versus the effective coupling
constant. The solid line is obtained from the exactly solved model
of Lieb--Linger gas at zero temperature. The symbols show the
experimental data.\cite{55}}
\end{center}
\vspace*{2mm}

In fact, many experiments have successfully demonstrated the confinements of atoms to one dimension by optical waveguides.
Another particularly interesting example involves the measurement of
photoassociation rates in one-dimensional Bose gases of ${}^{87}$Rb
atoms to determine the local pair correlation function $g^{(2)}(0)$
over a range of interaction strengths, see Fig.~\fref{f13}.
This experiment provides a direct observation of the fermionization of bosons with increasing interaction strength.
It sheds light on the phase coherence
behavior.\cite{28,33,41,42,80} At zero temperature, the local pair
correlation is $g^{(2)}(0) \sim 1$ for the weakly interacting Bose
gas and $g^{(2)}(0)\to 0$ as the system  enters  into  the
Tonks--Girardeau regime.

As discussed in previous sections, the finite-temperature problem of
the Lieb--Liniger Bose gas was solved by Yang and Yang in
1969.\cite{3}
It turns out that the Yang--Yang thermodynamic equation is an elegant way to analytically access the thermodynamics, quantum fluctions, and quantum criticality.
The Yang--Yang thermodynamics have been confirmed in the recent
experiments  through  various thermodynamical
properties\cite{61,62} and  quantum fluctuations.\cite{64,65,66}
A typical example is the measurement of the Yang--Yang
thermodynamics on the atom chip,  see Fig.~\fref{f14}.

\vspace*{2mm}

\begin{center}
\fl{f14}\includegraphics[width=\linewidth]{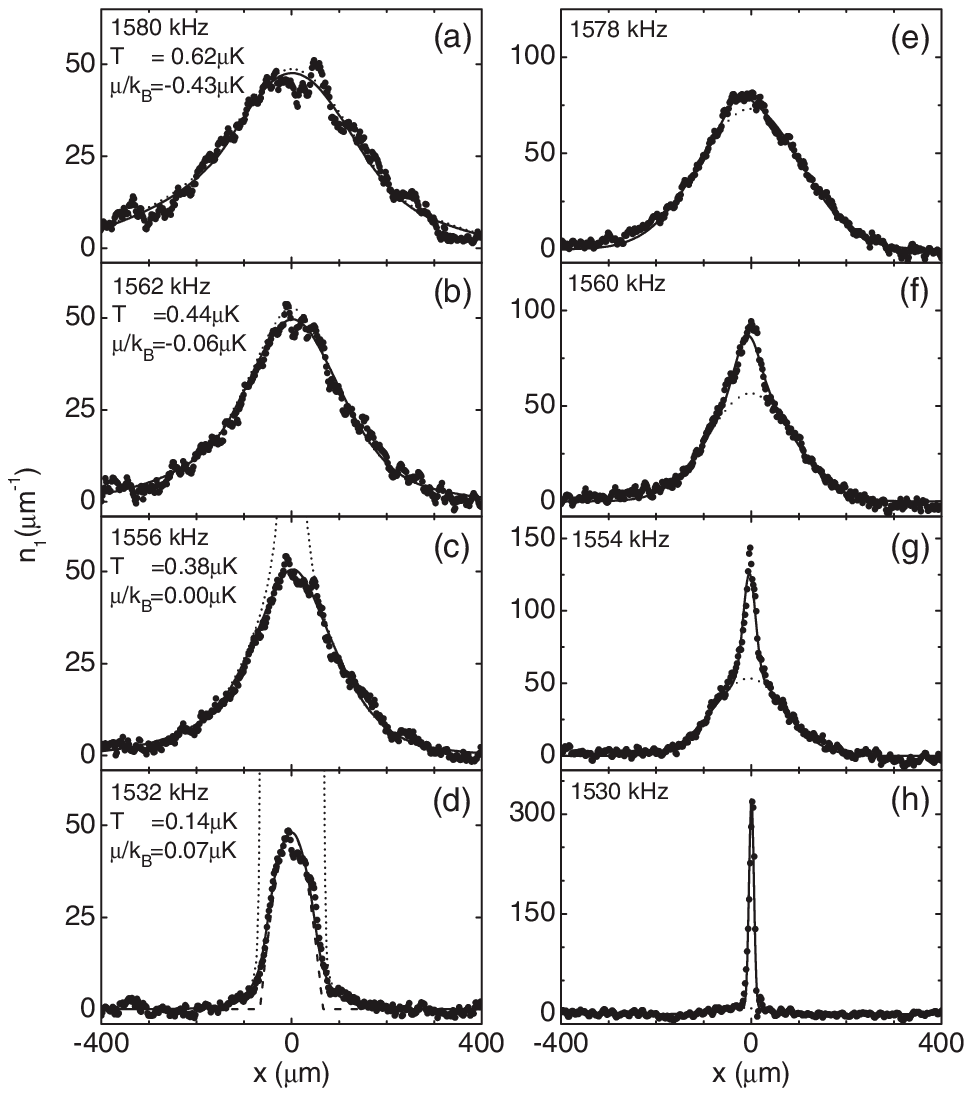}\\[5pt]
\parbox[c]{8.0cm}{\footnotesize{\bf Fig.~14.}
The {\em in situ} axial density profiles for the weakly interaction Bose
gas of ${}^{87}$Rb atoms at different temperatures. The solid lines
show the result obtained from the Yang--Yang equations. The values
of the chemical potentials are indicated by the harmonically
trapping potentials. The experimental data show a good agreement
with the theoretical prediction from the Yang--Yang
equation.\cite{54}}
\end{center}
\vspace*{2mm}

Moreover, recent experimental simulations with ultracold atoms
provide promising opportunities to test quantum dynamics of
many-body systems.
In particular, nonequilibrium  evolution of an isolated system involves transport and quench dynamics beyond the usual thermal Gibbs mechanism where the ground state and low lying excitation play an essential role.
The experimental study\cite{69,70} of thermalization of 1D ensemble
of cold atoms has led to significant developments in this
field.\cite{81,82}
In these experiments, it was demonstrated that quenching the dynamics into the isolated systems can lead to non-thermal distributions if conserved laws exist.
So far  a  generalized Gibbs ensemble is believed to present the non-thermal distributions in the isolated systems with conserved laws.
The many-body density matrix is written as
\begin{eqnarray}\el{87}
\hat{\rho} =\frac{1}{Z}\exp\left(-\sum_{m}\lambda_m \hat{ \cal{I} }_m \right)
\end{eqnarray}
in terms of conserved quantities $\hat{ \cal{I} }_m$.~Here
$Z={\rm Tr}\exp\left(-\sum_{m}\lambda_m \hat{ \cal{I} }_m \right)$ is the
partition function.
The Lagrange multipliers $\lambda_m$ acting for maximization of the entropy are determined by the associated conserved laws.
Figure~\fref{f15} shows the quantum Newton's cradle that provides
an insightful signature of such a generalized Gibbs ensemble.
It shows that the 1D systems with many-conserved laws do not approach the thermal equilibrium.

\vspace*{2mm}

\begin{center}
\fl{f15}\includegraphics[width=\linewidth]{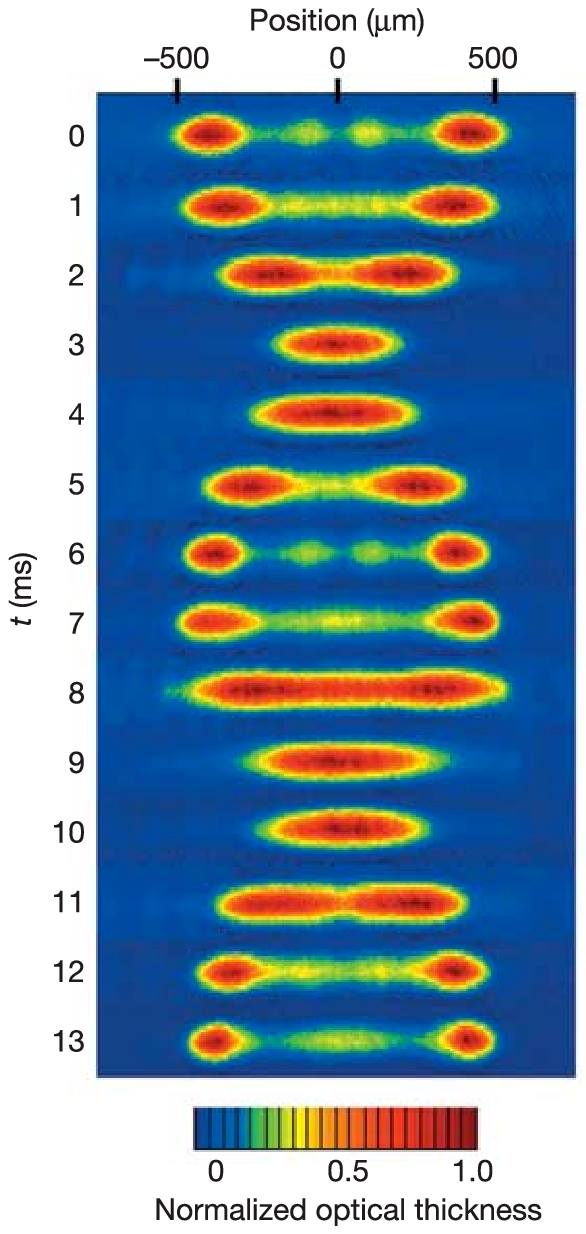}\\[5pt]
\parbox[c]{8.7cm}{\footnotesize{\bf Fig.~15.}
The time series of absorption images of the first oscillation cycle
for initial average peak coupling strength $r_0=1$. The two groups
of the cold atoms were confined in one dimension and initially
separated by grating pulses. They evolved from time to time and
collided twice in the centre  of  the harmonic 1D trap in each full
cycle. The oscillations in 1D Bose gas last for a long time, even
without approaching equilibrium.\cite{69}}
\end{center}

Recently, Langen {\em et al.}\cite{75} showed that a degenerate 1D
Bose gas relaxes to a state that can be described by such a
generalized Gibbs ensemble.
By splitting a 1D Bose gas into two halves, they prepared a non-equilibrium system of ${}^{87}$Rb atoms trapped in an atomic chip.
They measured the local relative phase profile $\varphi(z)$ between the two halves.
It was shown that most of the experimentally reachable initial states evolve in time into the steady states which can be determined within a reasonable precision by far less than $N$ Lagrange multipliers.
It was  particularly interesting to see that the experimental data of the  reduced $\chi^2$ values   can be well fitted with about 10 modes although there exist a much larger number of conserved quantities in the system.
This research further opens the study of the generalized Gibbs ensemble for the quantum systems out of equilibrium.

\section{Outlook}

We have introduced a fundamental understanding of many-body phenomena in the Lieb--Liniger model.
  The exact results for various physical properties of the Lieb--Liniger  model at $T=0$ and at finite temperature were obtained  by using the Bethe ansatz equations.
  In particular, we have presented a precise understanding of the excitation modes, Luttinger liquid, quantum statistics,  quantum criticality, correlations, and dynamics in the context  of Bethe ansatz.
In fact,  there have been great developments  in the study of the
Lieb--Liniger model in Refs.~\cite{11,15,36} via various methods,
such as field theory methods,\cite{33,83} Luttinger liquid theory
and bosonization, etc.\cite{35,84} It was shown that the  repulsive
Lieb-Liniger Bose gas  can be obtained as the nonrelativistic limit
of the sinh-Gordon model.\cite{85,86,87} Moreover, the study of the
non-thermal distributions for the isolated systems with many
conserved laws has attracted much attention.  In this scenario, the
generalized Gibbs ensemble\cite{82,88,89,90,91}  has been used  to
study the thermalization of the isolated systems. More recently,
there has been growing interest on quench dynamics in terms of the
generalized Gibbs Ensemble.\cite{92,93,94,95,96,97}
 This research has been  becoming a new frontier in cold atoms and condensed matter physics.
 It turns out that the integrable systems of this kind thus provide a promising  platform to advance the basic understanding of new quantum effects in many-body physics, such as few-body problems, universal thermodynamics, universal contact, quench dynamics, and correlation functions.
These studies will further place mathematical theories of exactly solvable models into the laboratory for a wide range of physical phenomena.

\section*{Acknowledgment}

This brief introduction to the Bethe ansatz solvable model is based on some informal lectures delivered by the author Guan X W at the University of Science and Technology of China, and Institute for Advanced Study at Tsinghua University.

The author Guan X W thanks Murray T Batchelor,  You-Jin Deng, Hui
Zhai, Zhen-Sheng Yuan, Fei Zhou, and Qi Zhou for helpful discussion.
He acknowledges the Department of Physics, Chinese University of
Hong Kong for their kind hospitality.


\begin{thebibliography}{99}
\bibitem {72} The experiments are chosen for the most relative to the Lieb--Liniger model, and by no means exhaustive.
\end{thebibliography}
\end{document}